\documentclass[paper]{JHEP}

\usepackage{amssymb}
\usepackage{bbm}
\usepackage{amsmath}
\usepackage{makeidx}
\usepackage{amsfonts}
\usepackage{amsthm}
\usepackage{longtable}
\usepackage[dvips]{graphicx}
\usepackage[title,titletoc,toc,page]{appendix}
\usepackage{pstricks}
\newcommand{\e}{\textrm{e}}                         % exponent e, e=2.71828.....
\newcommand{\f}[1]{\footnotesize{$ #1 $}}           % to put greek letter next to pspicture-objects
\renewcommand{\det}{\textrm{det}}                   % determinant function; straight up, i.e text-style
\renewcommand{\d}{\textrm{d}}                       % differential d; straight up
\newcommand{\Real}{\textrm{I\!R}}                   % real numbers
\newcommand{\Tr}{\textrm{Tr}}                       % trace ... textstyle
\newcommand{\ad}{\textrm{ad}}                       % lie algebra adjoint action; textstyle
\newcommand{\id}{\textrm{id}}                       % identity operator
\newcommand{\mf}[1]{\mathfrak{#1}}                  % short for mathfrak-style
\newcommand{\ft}[1]{\footnotesize{#1}}              % text footnotesize
\title{Coset Symmetries in Dimensionally Reduced Heterotic Supergravity}

\author{D.~B.~Westra, W.~Chemissany\\
Institute for Theoretical Physics\\
   Nijenborgh 4, 9747 AG Groningen,
     The Netherlands\\E-mail: \email{d.b.westra@rug.nl,
     w.chemissany@rug.nl}}

\preprint{\today\\UG-05-06\\ \hepth{0510137}}

\abstract{We investigate the coset structures which appear in the
dimensional reduction of supergravity theories. Especially we
investigate how to recognize the global symmetry groups if the
coset is non-split. As an example we apply our analysis to the
theories emanating from the dimensional reduction of Heterotic
supergravity.}

\keywords{Global Symmetries, Supergravity Models}

\begin{document}
\psset{unit=0.88cm}

%%%%%%%%%%%%%%%%%%%%%%%%%%%%%%%%%%%%%%%%%%%%%%%%%%%%%%%%%%%%%%%%%%%%%%%%%%%%%%%%%%%%%%%%%%
\section{Introduction}

Supergravity theories provide a useful playground for probing
string theory physics (see e.g. \cite{strings} and references
therein), since they are on the one hand a limit of string theory,
but on the other hand they are classical field theories, admitting
a more simpler analysis than string theory.

The scenario of dimensional reduction provides us with a setting
in which an effective four dimensional theory can be obtained from
the ten dimensional supergravity theories, therefore bringing four
dimensional physics in contact with string theory. The scalars in
supergravity theories often parameterize cosets $G/K$ where $K$ is
the maximal compact subgroup of the global symmetry group $G$.
This paper is about the cosets $G/K$. The global symmetry group
$G$ is related to the U-duality group, which contains the S- and
T-duality of string theories\cite{Duff94,Hull94,italy}.

In maximal supergravities it was shown \cite{LuPop,LuPop2} that in
a circle by circle reduction all scalars appeared in
upper-triangular matrices parameterizing the solvable positive
root subalgebra of the global symmetry group. Recognizing the
group $G$ is greatly facilitated by the fact that the Lie algebra
in these reductions of maximal supergravities is a split real
form, because then all dilaton coupling vectors can be identified
with positive roots and a Dynkin diagram can be drawn.

For non-maximal supergravities the Lie algebra of the global
symmetry group can be a non-split real form. In this paper we will
address the question of how to recognize the global symmetry group
parameterized by the scalars in the case where the Lie algebra is
not a split real form. We find that the identification of roots is
replaced by the identification of restricted roots and that the
Dynkin diagram is that of the restricted root system. Together
with the multiplicities of the restricted roots this fixes
uniquely the global symmetry group. Theories where scalars
parameterize cosets $G/K$ where $G$ is a non-split real form have
been studied before, see e.g. \cite{oxi,Yilmaz}, but in this work
the groups $G$ and $K$ were known beforehand. We present a
technique to find the groups $G$ and $K$ in theories obtained from
a dimensional reduction.

In this paper we will focus on the dimensional reduction of
Heterotic supergravity as a relevant example where non-split real
forms arise. The result is known\cite{MS,LuPop3,KM} but in this
paper we present a method which can be used for any supergravity
theory and which gives more insight in how the cosets appear in
supergravity theories.

The paper is organized as follows; in section \ref{sectionmethod}
we perform the dimensional reduction of Heterotic supergravity to
outline the method of dimensional reduction. In section
\ref{secalgebra} we show the relation between restricted roots and
scalar coset Lagrangians; appendix \ref{appendixLie1} is a quick
reference for Lie algebraic concepts and explains our notational
conventions on Lie algebras. In section \ref{secanalysis} we
analyze the reduction of the higher dimensional symmetries of
Heterotic supergravity and show that they give the same symmetry
group as obtained by the method of section \ref{secalgebra}. In
section \ref{secmaximal} we discuss the concept of a maximal
scalar manifold and show for Heterotic supergravities how field
dualizations give symmetry enhancements. In section
\ref{conclusions} we draw some conclusions.

%%%%%%%%%%%%%%%%%%%%%%%%%%%%%%%%%%%%%%%%%%%%%%%%%%%%%%%%%%%%%%%%%%%%%%%%%%%%%%%%%%%%%%%%%%
%%%%%%%%%%%%%%%%%%%%%%%%%%%%%%%%%%%%%%%%%%%%%%%%%%%%%%%%%%%%%%%%%%%%%%%%%%%%%%%%%%%%%%%%%%
\section{The Dimensional Reduction Method}\label{sectionmethod}
The method of dimensional reduction used in this paper is similar
to that of \cite{LuPop,LuPop2}. Actually a Kaluza-Klein circle by
circle reduction is performed; the total number of circles reduced
upon is called $\d$ and the dimension of the field theory is
$D=10-\d$. Since the global symmetry group manifests itself
already on the bosons in the theory and since the Kaluza-Klein
procedure does not break supersymmetry, we will only be concerned
with the bosonic field content. The fields are: the metric
$g_{\mu\nu}$, the Yang-Mills field $A_{\mu}$ in some
representation of either $E_8\times E_8$ or $SO(32)$, the dilaton
$\Phi_0$ and the Kalb-Ramond gauge potentials $B_{\mu\nu}$. As
usual we will restrict ourselves to the abelian subalgebra of the
Yang-Mills sector and therefore only 16 gauge bosons remain, but
we will not restrict ourselves to this number 16 and just assume
the existence of $N$
abelian gauge bosons.\\
The action can be written in Einstein frame as
\begin{equation}
S = \int_{\mathcal{M}_{10}}d^{10} \, x \, e\left( R -
\frac{1}{2}(\partial \Phi_0)^2 -\frac{1}{12}\e ^{-\Phi_0}H^2
-\frac{1}{4}\e^{-\tfrac{1}{2}\Phi_0}\sum_{I=1}^{16}
F^{I}_{\mu\nu}F^{I\mu\nu} \right),
\end{equation}
where $e=\det(e_{\mu}^{a})$ and $F^I=\d A^I$. The field strength
$H$ contains the Yang-Mills Chern-Simons term: $ H=\d B -
\sum_{I=1}^{N}\tfrac{1}{2} A^I\wedge F^I$.

\subsection{Reduction over one circle}
To obtain the $(10-\d)$-dimensional theory, we reduce $\d$ times
over a circle. In going from $D+1$ to $D$ dimensions we write
\begin{equation}
\d s^{2}_{D+1} = \e^{2\alpha\varphi}\d s^{2}_{D} +
\e^{-2\alpha(D-2)\varphi}(dz+ V_{\mu}\d x^\mu )^2 ,
\end{equation}
where the number $\alpha$ is given by
\begin{equation}
\alpha = \sqrt{\frac{1}{2(D-1)(D-2)}}\equiv \tfrac{1}{2}s ,
\end{equation}
and where $z$ is the coordinate over the circle. All fields are
independent of $z$. The Kaluza-Klein vector is denoted by
$V_{\mu}$ and reducing over more then one dimension will result in
a set of Kaluza-Klein vectors $V_{\mu}^{i}$. To obtain the action
we use the following rules for the dimensional reduction of the
Ricci scalar and an $n+1$-form field strength $F_{n+1}$ at every
step in going from $D+1$ to $D$ dimensions:
\begin{subequations}
\begin{align}
\sqrt{- \hat g}\hat R &= \sqrt{-g}\left( R - \tfrac{1}{2}(\partial
\varphi)^2-\tfrac{1}{4}\e^{-(D-1)s\varphi } F(V)^2 \right) \\
\tfrac{1}{2(n+1)!}\sqrt{-\hat g} \hat F_{(n+1)}^{2} &=
\tfrac{1}{2(n+1)!}\sqrt{-g} F^{2}_{(n+1)}\e^{-ns\varphi} +
\sqrt{-g}\tfrac{1}{2\, n!}F_{(n)}^{2}\e^{(D-n-1)s\varphi}
\label{form-reduction}.
\end{align}
\end{subequations}
This is enough to determine the action in any dimension $D$. The
fields descending from the ten dimensional metric are: the metric
$g_{\mu\nu}$, the dilatons $\varphi_i$, the Kaluza-Klein vectors
$V_{\mu}^{i}$ and the axions, which actually descend from the
Kaluza-Klein vectors, $A_{ij}$ and are thus defined only for
$i<j$. The ten dimensional Kalb-Ramond gauge potential gives rise
to a two-form $B_{\mu\nu}$, vectors $B_{\mu i}$ and scalars
$B_{ij}$. The ten dimensional Yang-Mills field gives rise to a
Yang-Mills vector $A_{\mu}^{I}$ and scalars $A^{I}_{i}$.

\subsection{Reduction over $\d$ circles}
The circle by circle reduction from 10 to $D$ dimensions can be
seen as a dimensional reduction over a $\d$-torus with the
following metric Ansatz:
\begin{equation}
\d s_{10}^2 = \e^{\tfrac{1}{2}\vec g\cdot \vec \varphi}\d
s_{D}^{2} + \sum_{i=1}^{\d} \e^{2\vec \gamma_i \cdot
\vec\varphi}(h^i)^2, \ \d+D=10,
\end{equation}
where the toroidal coordinates are denoted $z^i$ and
\begin{subequations}
\begin{align}
\vec \varphi & = (\varphi_1,\ldots,\varphi_\d ),\quad \vec g = 2(s_1,s_2,\ldots,s_{\d}),\\
\vec \gamma_i &= \frac{1}{4}\vec g - \frac{1}{2}\vec f_i, \quad\,
\vec f_i =
(\underbrace{0,\ldots,0}_{i-1},(9-i)s_i,s_{i+1},\ldots,s_{\d})
,\\
h^i &= dz_i + V_{\mu}^{i}\d x^\mu + \sum_{i<j\leq n}A_{ij}dz_j
\label{hi-definition}.
\end{align}
\end{subequations}
Equation \ref{hi-definition} can be inverted to give
\begin{equation}
dz^i = \Gamma_{ij}(h^j - V^j)
\end{equation}
where
\begin{equation}
\Gamma_{ij}  = \left(\sum_{p=0}^{\infty}(-A)^{p} \right)_{ij} =
\delta_{ij} - A_{ij} + A_{im}A_{mj} - \ldots,
\end{equation}
and it should be noted that due to fact that $A_{ij}$ is only
defined for $i<j$ the matrix $(A)_{ij} = A_{ij}$ is
upper-triangular and hence the series for $\Gamma_{ij}$ is finite.
Using a hat for generic higher dimensional fields and the
coordinate split $\hat x = (x^\mu, z^i)$ the Kaluza-Klein Ansatz
for the Yang-Mills gauge potential is
\begin{equation}
\hat A^I (\hat x) = A^{I}_{\mu}(x)\d x^\mu + A^{I}_{i}(x)dz^i =
A^I + A^{I}_{i}dz^i,
\end{equation}
and hence \footnote{The wedge symbols $\wedge$ will be omitted
where possible.}
\begin{subequations}
\begin{align}
\hat F^I & = \d A^I + \d A^{I}_{i}dz^i = F^I + F^{I}_{i}h^i \\
F^{I}_{i} &=(\d A_{m}^{I})\Gamma_{mi} \\
F^I &= \d A^I -F^{I}_{i}V^i .
\end{align}
\end{subequations}
For the Kalb-Ramond field we proceed analogously by expanding the
two-form field as
\begin{equation}
\hat B = B + B_idz^i + \tfrac{1}{2}B_{ij}dz^idz^j,
\end{equation}
and then calculating $\d \hat B$ and expressing this in the
$h^i$-basis. By incorporating the Chern-Simons terms in a similar
fashion one obtains \footnote{Any repeated index, whether in an
up-down combination or not, is summed over unless otherwise
specified.}
\begin{subequations}
\begin{align}
\hat H &= H + H_{i}h^i +\tfrac{1}{2}H_{ij}h^ih^j ,\\
H & = \d B -(\d B_i)\Gamma_{ij}V^j +\tfrac{1}{2}(\d
B_{ij})\Gamma_{im}\Gamma_{jn}V^mV^n -\tfrac{1}{2}(A^I -A^I_{i}\Gamma_{ij}V^j) F^I ,\\
H_n & = (\d B_i)\Gamma_{in} - (\d B_{ij})\Gamma_{im}\Gamma_{jn}V^m
-\tfrac{1}{2} (A^I -A^I_{i}\Gamma_{ij}V^j)F^{I}_{n} -\tfrac{1}{2}A^{I}_{n}F^I ,\\
H_{mn} & = (\d B_{ij})\Gamma_{im}\Gamma_{jn} + \tfrac{1}{2}\left(
A^{I}_{p}\Gamma_{pm} F^{I}_{n} - A^{I}_{p}\Gamma_{pn} F^{I}_{m}
\right) .
\end{align}
\end{subequations}
In a analogous way one obtains for the Kaluza-Klein field
strengths:
\begin{subequations}
\begin{align}
\hat F^i & = F^i + F_{ij}h^j ,\\
F_{ij} &= (\d A_{im} )\Gamma_{mj} ,\\
F^i & = \d V^i - (\d A_{im})\Gamma_{mn}V^n = \d V^i -F_{ij}V^j .
\end{align}
\end{subequations}
The dilatons from the reduction $\varphi_i$ have the usual field
strengths $\d \varphi_i$. In writing down the action it is
convenient to put all dilatons, both from the reduction and the
ten dimensional dilaton, into a $\d+1$-component vector, and hence
all dilaton couplings into $\d+1$-component `coupling vectors'.
The useful definitions are
\begin{equation}
\begin{array}{l@{\,\,\quad}l}
\vec \Phi = (\Phi_0,\varphi_1,\ldots,\varphi_\d) = (\Phi_0,\vec \varphi), & \vec F_i  = (0,\vec f_i ),\\
\vec B_{ij} = -\vec F_i +\vec F_j ,& \vec G  = (1,\vec g)  ,\\
\vec A_{i} = \vec F_i - \vec G ,& \vec B_i = -\vec F_i  ,\\
\vec A_{ij} = \vec F_i+\vec F_j -\vec G ,& \vec A  = -\vec G ,\\
C_{i} = \vec F_i -\tfrac{1}{2}\vec G, & \\
\end{array}
\end{equation}
of which some summation relations can be deduced
\begin{subequations}
\begin{align}
\vec A_{ij} + \vec B_{ik} &= \vec A_{ki},\ \ \vec A_{ij} + \vec
B_{jk} = \vec A_{ik} ,\\
\vec C_{i} + \vec C_j &= \vec A_{ij} , \ \ \vec C_j = \vec B_{ij}
+ \vec C_i, \ i<j,
\end{align}
\end{subequations}
and some inner product relations
\begin{subequations}
\begin{align}
\vec G \cdot \vec G & = \tfrac{8}{D-2}, \quad \vec F_i\cdot \vec
F_j = 2\delta_{ij}+\tfrac{2}{D-2}, \\
\vec F_i \cdot \vec G & = \tfrac{4}{D-2}, \quad \vec A_{i}\cdot
\vec G = -\tfrac{4}{D-2}, \\
\vec A_{ij}\cdot \vec G &= 0,\quad \vec B_{ij}\cdot \vec
B_{kl} = 2\delta_{ik}-2\delta_{il}-2\delta_{jk}+2\delta_{jl}, \\
\vec A_{ij}\cdot \vec B_{kl} & =
-2\delta_{ik}+2\delta_{il}-2\delta_{jk}+2\delta_{jl}, \\
\vec C_{i}\cdot \vec C_{j} &= 2\delta_{ij}.
\end{align}
\end{subequations}
The $D$-dimensional Lagrangian can be written as
\begin{subequations}
\begin{align}
\mathcal{L} & = \mathcal{L}_1 + \mathcal{L}_2 + \mathcal{L}_3 \label{L1},\\
e^{-1}\mathcal{L}_1& = R -\tfrac{1}{2} \partial_\mu \vec\Phi\cdot
\partial^\mu \vec \Phi-\tfrac{1}{2}\sum_{1\leq i<j\leq \d} (F_{ij})^2\e^{\vec
B_{ij} \cdot \vec \Phi} -\tfrac{1}{4} \sum_{i=1}^{\d}
(F_i)^2\e^{\vec B_i \cdot \vec \Phi}\\
e^{-1}\mathcal{L}_2 &= -\tfrac{1}{12}\e^{\vec A\cdot \vec \Phi}
H^2 -\tfrac{1}{4}\sum_{i=1}^{\d} \e^{\vec A_i \cdot \vec \Phi}
(H^i)^2 - \tfrac{1}{2} \sum_{1\leq i<j \leq \d}\e^{\vec
A_{ij}\cdot
\vec \Phi} (H_{ij})^2\\
e^{-1} \mathcal{L}_3 &= -\tfrac{1}{4} \e^{-\tfrac{1}{2}\vec G
\cdot \vec \Phi}\sum_{I=1}^{N} (F^I)^2
-\tfrac{1}{2}\sum_{i=1}^{\d}\sum_{I=1}^{N} \e^{\vec C_{i}\cdot
\vec \Phi } (F^{I}_{i})^2 .
\end{align}
\end{subequations}

%%%%%%%%%%%%%%%%%%%%%%%%%%%%%%%%%%%%%%%%%%%%%%%%%%%%%%%%%%%%%%%%%%%%%%%%%%%%%%%%%%%%%%%%%
%%%%%%%%%%%%%%%%%%%%%%%%%%%%%%%%%%%%%%%%%%%%%%%%%%%%%%%%%%%%%%%%%%%%%%%%%%%%%%%%%%%%%%%%%
\section{The Algebra from the Reduction}\label{secalgebra}

\subsection{Restricted Roots and Coset
Lagrangians}\label{subsecRes} In this section we will set out the
method for identifying the global symmetry group $G$ of the coset
$G/K$ which is parameterized by the scalars emanating from a
dimensional reduction. A summary of some Lie group theoretical
aspects is in appendix \ref{appendixLie1}. Every semi-simple real
Lie algebra $\mf{g}$ of a Lie group $G$ can be decomposed in a
compact subalgebra $\mf{k}$, a maximal abelian subalgebra $\mf{a}$
and a nilpotent subalgebra $\mf{n}$. This decomposition
$\mf{g}=\mf{k}\oplus\mf{a}\oplus\mf{n}$ is the Iwasawa
decomposition. The orthogonal component of $\mf{g}$ with respect
to the Cartan-Killing form $B(x,y)$ to $\mf{k}$ is denoted
$\mf{p}$ and we have
\begin{equation}
[\mathfrak{k}\,\mathfrak{k}]\subset \mathfrak{k}, \
[\mathfrak{k}\,\mathfrak{p}]\subset \mathfrak{p}, \
[\mathfrak{p}\,\mathfrak{p}]\subset \mathfrak{k}.
\label{commutatiecompact}
\end{equation}
If $K$ is a maximal compact subgroup of the Lie group $G$ with Lie
algebra $\mf{k}$, then we can describe the coset $G/K$ by
$\textrm{exp}(\mf{a}\oplus\mf{n})$\footnote{Of course only the
identity component of $G/K$ is parameterized in this way.}. The
scalar manifolds appearing in supergravities are in general
Riemannian globally symmetric spaces and can be described by
cosets of the form $G/K$, where $K$ is the maximal compact
subgroup of a semi-simple real Lie group $G$. These cosets $G/K$
are classified (see e.g. \cite{helgason}). The classification says
which maximal compact subalgebras $\mf{k}$ can be found in a real
semi-simple Lie algebra $\mf{g}$ and thus which real forms
$\mf{g}$ a complex Lie algebra $\mf{g}'\cong\mf{g}^\mathbbm{C}$
can have. With respect to the subalgebra $\mf{a}$ the Lie algebra
$\mf{g}$ can be decomposed into restricted root spaces
$\mf{g}_\lambda$:
\begin{equation}
\mf{g} = \mf{g}_0\oplus\bigoplus_{\lambda \in
\Sigma}\mf{g}_\lambda, \quad [Hx_\lambda]=\lambda(H)x_\lambda,
\forall H\in\mf{a},x_\lambda\in\mf{g}_\lambda ,
\end{equation}
where $\Sigma$ is the set of (nonzero) restricted roots. This
decomposition is analogous to the root decomposition with respect
to the Cartan subalgebra $\mf{h}^{\mathbbm{C}}$ of the
complexified algebra $\mf{g}^{\mathbbm{C}}$, but since the real
numbers do not form a closed field, $\mf{a}^{\mathbbm{C}}$ can not
be identified with $\mf{h}^{\mathbbm{C}}$, but only with a
subalgebra of the latter. Hence the name restricted root. The
restricted roots are linear real functionals on $\mf{a}$ and form
a root system \cite{helgason}, which can in fact be non-reduced,
i.e. if $\lambda \in \Sigma$ then the only multiples of $\lambda$
that can also be in $\Sigma$ are $\pm \lambda,\pm 2 \lambda, \pm
\tfrac{1}{2}\lambda$ (but if $\lambda, 2\lambda\in \Sigma$ then
$\tfrac{1}{2}\lambda\notin\Sigma$). Another major deviation from
the ordinary roots is that the dimension of the restricted root
spaces can exceed 1: $m_\lambda\equiv\dim \mf{g}_\lambda\geq 1$.
As with ordinary roots a set of simple restricted roots can be
defined and a Dynkin diagram can be drawn. The number of simple
restricted roots is called the rank and it equals the dimension of
$\mf{a}$. The multiplicities $m_{\lambda_i}$ and
$m_{2\lambda_i}=\dim \mf{g}_{2\lambda_i}$ of the simple restricted
roots $\lambda_i$ together with the restricted root diagram
uniquely fix the real form $\mf{g}$ of a complex Lie algebra
$\mf{g}'\cong \mf{g}^\mathbbm{C}$ and hence they fix the coset
$G/K$. In appendix \ref{tables} we list for all real non-compact
forms the restricted root diagram, the multiplicities of the
simple restricted roots and the Satake diagram (see appendix
\ref{appendixLie1} for some explanation).

As seen in the previous section, a circle by circle dimensional
reduction reveals the dilaton coupling vectors and these we will
identify with the set of positive restricted roots $\Sigma^+$. The
multiplicities are easily determined; they are just the number of
times a dilaton coupling with that restricted root occurs. A
restricted root Dynkin diagram can readily be drawn and thus the
corresponding coset can be read off from the tables in appendix
\ref{tables}. It only needs a proof that the scalars in the
Lagrangian really make up a coset Lagrangian. Therefore in the
remainder of this section we will sketch some aspects of coset
Lagrangians (see also \cite{oxi,Yilmaz,MS}).

The nilpotent subalgebra $\mf{n}$ is actually the subalgebra
\begin{equation}
\mf{n} = \bigoplus_{\lambda\in\Sigma^+} \mf{g}_\lambda .
\end{equation}
Let us introduce a basis $\left\{H_1,\ldots,H_l\right\}$ for
$\mf{a}$, where $l=\dim\mf{a}$ is the rank of the real form
$\mf{g}$ and let us fix a basis for $\mf{n}$ by the elements
$E_{\lambda}^{I}$, where for fixed $\lambda\in\Sigma^+$ the index
$I$ runs from 1 to $m_\lambda$. The coset $G/K$ can than be
described by scalars $\phi_i$, called dilatons, and by scalars
$A_{\lambda}^{I}$, called axions, through $\mathcal{V}
=\mathcal{V}_1 \mathcal{V}_2$ which is an element of $G$ and where
\begin{equation}
\mathcal{V}_1 =\exp{\tfrac{1}{2}\sum_{i=1}^{l}\phi_iH_i}, \,\quad
\mathcal{V}_2=\exp{\sum_{\lambda\in\Sigma^+}\sum_{I=1}^{m_\lambda}
A_{\lambda}^{I} E_{\lambda}^{I}}.
\end{equation}
As we will see later it is sometimes more convenient to
parameterize $\mathcal{V}_2$ slightly differently. We arrange the
dilatons $\phi_i$ in a vector $\vec \phi$ and similar for $H_i$.
For the restricted roots we define $\vec \lambda$ as the vector
with components $\lambda_i = \lambda(H_i)$. From $\mathcal{V}$ we
can compute the Lie algebra valued one-form\footnote{In deriving
these formula one uses the Baker-Campbell-Hausdorff formulae,
which can be found in appendix \ref{appendixBCH}. }
\begin{equation}
\d \mathcal{V} \mathcal{V}^{-1} = \tfrac{1}{2}  \d \vec\phi \cdot
\vec H + \sum_{\lambda \in \Sigma^+}\sum_{I=1}^{m_\lambda}\e^{\vec
\lambda\cdot \vec \phi} F_{\lambda}^{I}E_{\lambda}^{I} .
\end{equation}

With every real form $\mf{g}$ goes a Cartan involution $\theta$
(see appendix \ref{appendixLie1} but also \cite{helgason,knapp})
which is $+\id$ on the compact subalgebra $\mf{k}$. This Cartan
involution is used to define a generalized transpose $\#$ in the
(identity component of the) real group $G$ as follows: if $O\in G$
and $O=\exp{x}$ for some $x\in\mf{g}$, then $O^\# = \exp{-
\theta(x)}$. In fact if $U\in K\subset G$, then we have $U^\# =
U^{-1}$, which clarifies the name generalized transpose, since for
$SO(n)\ni O$ we have $O^{-1}=O^T$. A general scalar coset action
is of the form
\begin{equation}
S_{G/K} = \tfrac{1}{8}\int \, d^Dx\, e\Tr \left(
\partial \mathcal{M}\partial \mathcal{M}^{-1} \right) .
\end{equation}
where the trace is in some representation and
$\mathcal{M}=\mathcal{V}^\# \mathcal{V}$. Though $\mathcal{M}$ is
in a representation of the group, the trace in the action is
actually in a Lie algebra representation and using that $\theta$
is an automorphism one can show that the action can be written as:
\begin{equation}
\begin{split}
S_{G/K} &= -\tfrac{1}{4}\int \, \d^Dx e \left( \Tr( \partial
\mathcal{V} \mathcal{V}^{-1} \partial \mathcal{V} \mathcal{V}^{-1}
) + \Tr (
\partial \mathcal{V} \mathcal{V}^{-1} (\partial \mathcal{V} \mathcal{V}^{-1})^\# ) \right)\\ &=
-\tfrac{1}{2}\int \, \d^Dx e  \Tr( \partial \mathcal{V}
\mathcal{V}^{-1} \mathbbm{P}
\partial \mathcal{V} \mathcal{V}^{-1} ) ,\label{scalarcoset}
\end{split}
\end{equation}
where $\mathbbm{P}: \mathfrak{g}\rightarrow \mathfrak{g}$ denotes
the projection operator defined by
\begin{equation}
\mathbbm{P}: x\mapsto \tfrac{1}{2}(\mathbbm{1} - \theta )x .
\end{equation}
Hence $\mathbbm{P}$ is the identity on the non-compact part and
zero on the compact part; it projects out the compact part. This
is quite general for scalar coset Lagrangians; one starts with a
representative $V$ of the group $G$ parameterized by scalars and
writes
\begin{equation}
\d V V^{-1} = Q + P, \ Q\in\mathfrak{k},P\in\mathfrak{p} .
\end{equation}
Under a global $G$ transformation $V\mapsto V M, \, M\in G$ the
forms $Q$ and $P$ are invariant. From the relations
\ref{commutatiecompact} one finds that under a transformation
$V\mapsto OV, \, O\in K$ we have
\begin{equation}
Q\mapsto \d O O^{-1} + OQO^{-1}, \ P\mapsto OPO^{-1} ,
\end{equation}
and thus $Q$ is like a gauge field and $P$ transforms covariant.
Hence we can form the Lagrangian
\begin{equation}
L_{G/K} = -\tfrac{1}{2}\Tr ( P_{\mu}P^{\mu} ),
\end{equation}
which is precisely the same is
\begin{equation}
-\tfrac{1}{2}\Tr ( \mathbbm{P}\d V V^{-1} \mathbbm{P} \d V V^{-1})
=-\tfrac{1}{2}\Tr (\d V V^{-1} \mathbbm{P} \d V V^{-1}),
\end{equation}
where the latter equality follows from $\Tr(xy) \sim \Tr_{\ad}
(\ad x \ad y)= 0,\,\forall x\in\mathfrak{k},y\in\mathfrak{p}$.

Another approach for coset Lagrangian is to start with the
Lagrangian
\begin{equation}
L_{G/K}' = -\tfrac{1}{2} \Tr (D_{\mu}V V^{-1} D^{\mu}V V^{-1} ),
\end{equation}
where the covariant derivative $D$ contains a gauge field $A_\mu$
taking values in $\mathfrak{k}$ and appears algebraically in the
action. The gauge field can be eliminated by its equation of
motion $\Tr (A_{\mu} D^{\mu}V V^{-1})=0$, which precisely means
that $D_{\mu} V V^{-1}\in \mathfrak{p}$. Hence the gauge field
cancels the compact part in $\partial_\mu V V^{-1}$ giving thus
the same Lagrangian: $L_{G/K}=L_{G/K}'$. If the scalar sector in
the Lagrangian obtained by dimensional reduction matches the
action \ref{scalarcoset} for the appropriate $G/K$, then indeed
the scalars from the reduction parameterize the coset $G/K$. In
the following section we will pursue this programme for the
dimensionally reduced Heterotic supergravity.

\subsection{Identifying Restricted Roots in the
Lagrangian}\label{subsecIdent} The restricted roots are easily
read off from the Lagrangian \ref{L1} to be $\vec B_{ij}$, $\vec
A_{ij}$ and $\vec C_i$ with multiplicities $m(\vec B_{ij}) =m
(\vec A_{ij})=1$ and $m(\vec C_i)=N$, while $m(2\vec C_i)=0$. The
simple restricted roots can be identified as follows:
$\lambda_{\d-i}\leftrightarrow\vec B_{i,i+1}$ and
$\lambda_d\leftrightarrow\vec C_1$ and hence the rank of the coset
is $\d$. The dilaton coupling vectors are $\d +1$ dimensional so
one direction in this vector space should be redundant. In fact in
\cite{LuPop3} it is shown that indeed for $\d<6$ one can split of
one component of the dilaton. We will take another approach; since
it is known that in four dimensions the global symmetry group can
be enlarged to an $SL(2;\Real)\times SO(6,6+N)$ we will embed the
symmetry group already in the larger group $SL(2;\Real)\times
SO(\d,\d+N)$ which has rank $\d+1$. It is easy to see that the
inner product in the restricted root space is proportional to the
inner product of the dilaton coupling vectors and thus the
restricted root Dynkin diagram is
\begin{center}
\begin{pspicture}(0,0.7)(6,1.5)
\pscircle(0.5,1){0.1}\qline(0.7,1)(1.2,1)%
\pscircle(1.4,1){0.1}\qline(1.6,1)(2.1,1)%
\psline[linestyle=dotted,dotsep=0.1,linewidth=0.05](2.2,1)(2.6,1)\qline(2.7,1)(3.2,1)%
\pscircle(3.4,1){0.1}\psline[doubleline=true,doublesep=0.03,arrows=->,arrowlength=0.6,arrowinset=0.55](3.6,1)(4.1,1)\pscircle(4.3,1){0.1}%
\put(0.4,1.2){\f{\lambda_1}}\put(1.3,1.2){\f{\lambda_2}}%
\put(3.3,1.2){\f{\lambda_{\d-1}}}\put(4.2,1.2){\f{\lambda_\d}}%
\end{pspicture}
\end{center}
and taking into account the multiplicities one can read off that
the coset should be of the type B\textit{I} or D\textit{I}. This
implies $G= SO(\d,\d+N)$, since the rank equals $l$ for both cases
and $N=2(r-l)$ if $N+2\d$ is even (type D\textit{I}) and
$N=2(r-l)+1$ if $N+2\d$ is odd (type B\textit{I}).

The rest of this section will thus be devoted to prove that indeed
the scalar part of the Lagrangian \ref{L1} is an
$SO(\d,\d+N)/SO(\d)\times SO(\d+N)$ coset Lagrangian. Some aspects
of the explicit representations come in handy at this point. With
every restricted root we can identify as many generators as the
multiplicity and also we assemble $\d+1$ non-compact\footnote{As
much as is possible we will omit the subjective non-compact; a
Cartan generator in this context is an element of $\mf{a}$.}
Cartan generators $H_a, \, 0\leq a \leq \d$ in a vector where the
$\mf{sl}(2;\Real)$ Cartan generator is embedded as a linear
combination. We therefore make the following identification:
\begin{subequations}
\begin{align}
\vec B_{ij} & \leftrightarrow E_{ij} , \ i<j, \ \quad\quad\quad \ \ [\vec H E_{ij} ] = \vec B_{ij} E_{ij}\label{B-vector},\\
\vec A_{ij} & \leftrightarrow R_{ij}=-R_{ji} ,\ \quad\quad\quad \ [\vec H R_{ij} ] = \vec A_{ij}R_{ij} \label{A-vector},\\
\vec C_{i} & \leftrightarrow Y_{iI} ,\, 1\leq I\leq N, \quad \quad
\quad [\vec H Y_{iI} ] = \vec C_{i}Y_{iI}.\label{C-vector}
\end{align}
\end{subequations}
The summation rules suggest that we take
\begin{equation}
\begin{array}{l@{\quad}l}
\left[E_{ij}E_{kl}\right] \sim  \delta_{jk}E_{il} -\delta_{il}E_{kj}, & \left[E_{ij} R_{kl}\right]\sim \delta_{ik}R_{jl} - \delta_{il}R_{jk}, \\
\left[E_{ij}Y_{kK} \right] \sim  \delta_{ik}Y_{jK}, & \left[Y_{iI}Y_{jJ} \right] \sim M_{IJ} R_{ij} ,\\
\left[Y_{iI}R_{kl}  \right] =0 ,& \left[R_{ij}R_{kl}\right]=0 ,\\
\end{array}
\end{equation}
where $M_{IJ}$ is an unknown matrix. Working out the Jacobi
equations fixes the proportionality constants but not $M_{IJ}$,
since this is related to a choice of basis in the subspace spanned
by the $Y_{iI}$. Using the vector representation with the basis as
in appendix \ref{appendixLie2} and working out the commutation
relations gives $M_{IJ}=\delta_{IJ}$. Hence we have:
\begin{equation}
\begin{array}{l@{\quad}l}
\left[E_{ij}E_{kl}\right] =  \delta_{jk}E_{il} -\delta_{il}E_{kj},& \left[E_{ij} R_{kl}\right]=- \delta_{ik}R_{jl} + \delta_{il}R_{jk} ,\\
\left[E_{ij}Y_{kK} \right] = -  \delta_{ik}Y_{jK} ,& \left[Y_{iI}Y_{jJ} \right] = \delta_{IJ} R_{ij},\\
\left[Y_{iI}R_{kl}  \right] =0 ,&\left[R_{ij}R_{kl}\right]=0.
\end{array}\label{commutator-i}
\end{equation}
The embedding in $SL(2;\Real)\times SO(\d,\d+N)$ is done by
embedding the vector representation $(2\d+N)\times (2\d+N)$
matrices $X$ (see appendix \ref{appendixLie2}) of the positive
restricted root vectors of $\mf{so}(\d,\d+N)$ into
$(2+2\d+N)\times (2+2\d +N)$ matrices as follows
\begin{equation}
X\hookrightarrow \begin{pmatrix} 0_{2\times 2} & 0_{2\times (2\d + N)} \\ 0_{(2\d +N)\times 2} & X \\
\end{pmatrix},
\end{equation}
and the Cartan generators are diagonal in this basis where the
upper left block is proportional to the third Pauli matrix. We
introduce for convenience the vector $\vec C_0 = \tfrac{1}{2}
\sqrt{D-2} \vec G$ to get
\begin{equation}
\vec C_a \cdot \vec C_b = 2\delta_{ab}, \quad 0\leq a,b \leq
\d;\quad \sum_{a=0}^{\d} (\vec C_a)_m (\vec C_a)_n =
2\delta_{mn}.\label{ortho}
\end{equation}
Inspection of the commutation relations \ref{commutator-i} and
using the explicit vector representation we see that the Cartan
generators $H_a$ are diagonal with diagonal values
\begin{equation}
\begin{split}
H_a &= diag ( (H_a)_0,-(H_a)_0,(H_a)_1,\ldots,(H_a)_\d,
-(H_a)_1,\ldots,-(H_a)_\d,\underbrace{0,\ldots, 0}_{N \textrm{ times}}),\\
&(H_m)_a = (-\vec C_a)_m, \, 1\leq a,m \leq \d .
\end{split}
\end{equation}
This implies that in the vector representation the traces are
\begin{equation}
\begin{split}
\Tr(H_aH_b) &= 2 \sum_{c=0}^{\d}(\vec C_c)_a (\vec C_c)_b =
4\delta_{ab},\\
\Tr ( E_{ij}E_{kl}^{T} ) &=    2\delta_{ik}\delta_{jl},\\
\Tr(R_{ij} R_{mn}^{T}) &= 2\delta_{im}\delta_{jn},
\end{split}
\end{equation}
and the others are zero. The traces in two different
representations of a simple Lie algebra are proportional and thus
for a semi-simple Lie algebra of two simple factors, there are two
proportionality constants since both factors can be weighted
differently. In our case we have fixed $\vec C_0$ and therefore we
have fixed one more constant and this choice gives the right coset
Lagrangian. The coset Lagrangian can be constructed using
\begin{subequations}
\begin{align}
\mathcal{V}_1& = \textrm{exp}\left( \tfrac{1}{2} \vec \Phi \cdot
\vec H
\right) ,\\
\mathcal{V}_2&= \cdots U_{24}U_{23} \cdots U_{14}U_{13}U_{12}, \
U_{ij}=\textrm{exp}\left( A_{ij}E_{ij} \right) \textrm{
no sum},\\
\mathcal{V}_3 &= \textrm{exp} \left( \sum_{i<j}
B_{ij}R_{ij}\right)\\
\Omega&= \textrm{exp}\left( \omega \right), \ \omega=
\sum_{iI}A_{iI}Y_{iI},\\
\mathcal{V}&= \mathcal{V}_1\mathcal{V}_2\mathcal{V}_3 \Omega,
\end{align}
\end{subequations}
where $A_{ij}$, $B_{ij}$ and $A_{iI}$ are the axions and
$\vec\Phi$ are the dilatons. Using some tricks as explained in
appendix \ref{appendixBCH} one finds:
\begin{subequations}
\begin{align}
d\mathcal{V}_1 \mathcal{V}_1{}^{-1} &= \tfrac{1}{2}\d \vec \Phi \cdot \vec H ,\\
\mathcal{V}_1 \d \mathcal{V}_2 \mathcal{V}_2{}^{-1}
\mathcal{V}_1{}^{-1} &=\sum_{i<j} F_{ij} \e^{\tfrac{1}{2}\vec
\Phi\cdot \vec
B_{ij}} E_{ij}\label{trick1},\\
\mathcal{V}_1\mathcal{V}_2 \d \mathcal{V}_3
\mathcal{V}_3{}^{-1}\mathcal{V}_2{}^{-1}\mathcal{V}_1{}^{-1}
&=\sum_{i<j}\e^{\tfrac{1}{2}\vec A_{ij} \cdot \vec \Phi} \d
B_{mn}\Gamma_{mi}\Gamma_{nj},\\
\mathcal{V}_1\mathcal{V}_2\mathcal{V}_3 d\Omega \Omega^{-1}\left(
\mathcal{V}_1 \mathcal{V}_2\mathcal{V}_3 \right)^{-1} &= \sum_{Ii}
\e^{\tfrac{1}{2}\vec C_{iI}\cdot \vec \Phi } F_{i}^{I}Y_{Ii}
+\tfrac{1}{2}\sum_{i,j}\e^{\tfrac{1}{2} \vec A_{ij} \cdot \vec
\Phi}A^{I}_{m}\Gamma_{mi}F^{I}_{j}R_{ij} \label{trick3}.
\end{align}
\end{subequations}
Using that the action of $\theta$ in the vector representation
becomes $\theta(x)= - x^T$ and using the properties of the traces
one then indeed finds that the coset construction based on the
commutation rules of the restricted root generators of
$\mf{so}(\d,\d+N)$ indeed gives the coset scalar Lagrangian as it
appears in the action \ref{L1}. Though we embedded the global
symmetry group into $SL(2;\Real)\times SO(\d,\d+N)$, $SL(2;\Real)$
is not a part of the symmetry group if $D>4$. This can be seen
from the fact that there is  no restricted root for $SL(2;\Real)$.
This will change when the two-form can be dualized to a scalar;
this can be done in four dimensions.

%%%%%%%%%%%%%%%%%%%%%%%%%%%%%%%%%%%%%%%%%%%%%%%%%%%%%%%%%%%%%%%%%%%%%%%%%%%%%%%%%%%%%%%%%
%%%%%%%%%%%%%%%%%%%%%%%%%%%%%%%%%%%%%%%%%%%%%%%%%%%%%%%%%%%%%%%%%%%%%%%%%%%%%%%%%%%%%%%%%
\section{Analysis of Dimensionally Reduced
Symmetries}\label{secanalysis} The fields appearing in the
Lagrangian \ref{L1} do not all have the `right' transformation
properties, e.g. the Kaluza-Klein $U(1)$-gauge transformation also
acts on other fields than the Kaluza-Klein vectors. The fields are
not `diagonalized' with respect to the symmetry transformations,
but a field redefinition can achieve this. An alternative but
equivalent approach is to slightly modify the reduction Ansatz in
such a way that the difference between world and tangent indices
is respected. Let us split up the world indices according to
$\hat\mu \rightarrow (\mu, \alpha)$ and the tangent space indices
according to $\hat a \rightarrow (a,i)$ and if more then one index
of a kind is needed, the order of the alphabet will be used. The
modification of the reduction Ansatz for the metric amounts to
\begin{equation}
h^i \rightarrow O^i{}_{\alpha}(\d z^\alpha +V_{\mu}^{\alpha}\d
x^\mu ), \ O^i{}_\alpha = \delta^{i}_{\alpha} + A^i{}_\alpha ,
\end{equation}
where $A^i{}_\alpha$ are the axions with the distinction between
flat and curved index expressed by using a different kind of
letter. The metric Ansatz can now be written in terms of vielbeins
by
\begin{equation}
\begin{split}
\hat e_{\mu}{}^{a} & \e^{\tfrac{1}{4}\vec g \cdot \vec
\varphi}e_{\mu}{}^a, \quad \hat e_{\mu}{}^i = \e^{\vec
\gamma_{i}\cdot \vec \varphi}O^i{}_\alpha V_{\mu}^{\alpha}, \\
\hat e_\alpha{}^i & = \e^{\vec \gamma_{i}\cdot \vec \varphi}
O^i{}_\alpha, \quad \hat e_{\mu}{}^i = 0 .
\end{split}
\end{equation}
The Ansatz of the fields will now be made in a tangent space basis
(see e.g. \cite{KM,rmd}). So for the Kalb-Ramond field we write:
\begin{equation}
\hat B = \tfrac{1}{2} B_{ab}e^a e^b + B_{a\alpha}e^af^\alpha
+\tfrac{1}{2}B_{\alpha\beta}f^\alpha f^\beta, \ f^\alpha \equiv \d
z^\alpha +V_{\mu}^{\alpha}\d x^\mu ,
\end{equation}
and for all other fields similar. Note that we still take
$A^i{}_{\alpha}$ to be upper triangular. The most general
diffeomorphism in ten dimensions compatible with the field
Ans\"atze is
\begin{equation}
\delta x^\mu = -\xi^\mu ( x ), \quad \delta  z^\alpha =
\Lambda^\alpha{}_{\beta}z^\beta + \xi^\alpha (x),
\end{equation}
where $\Lambda^\alpha{}_\beta$ is a constant matrix, hence a
member of $\mathfrak{gl}(\d;\Real)\sim \Real +
\mathfrak{sl}(\d;\Real)$ and from the discussion in \cite{LuPop}
we know that $\Real$-factor will combine with the higher
dimensional scaling symmetry to an internal abelian symmetry. We
will not be concerned with this symmetry. The
$\xi^\mu$-transformations give rise to lower dimensional general
coordinate transformation, while the $\xi^\alpha$-transformations
act only on the Kaluza-Klein vectors $V_{\mu}^{\alpha}$, which
transform under this transformation as $U(1)$-connections. Any
field of the form $C_{\mu_1\ldots\mu_p\alpha_1\ldots\alpha_q}$
transforms as a $q$-tensor under the
$\mathfrak{sl}(\d;\Real)$-transformations:
\begin{equation}
\delta C_{\mu_1\ldots\mu_p\alpha_1\ldots\alpha_q} =
C_{\mu_1\ldots\mu_p\beta\alpha_2\ldots\alpha_q}\Lambda^{\beta}{}_{\alpha_1}
+ \textrm{ other terms with $\alpha_1 \leftrightarrow \alpha_i$} .
\end{equation}
The Kaluza-Klein vector transforms under $\mf{sl}(\d;\Real)$ as
\begin{equation}
\delta_\Lambda V_{\mu}^{\alpha} = - \Lambda^{\alpha}{}_\beta
V_{\mu}^{\beta}.
\end{equation}
The Kalb-Ramond transformations act on the two-form field as
$\delta \hat B = \d \hat \Lambda^{(1)}$, and the only
transformation consistent with the reduction Ansatz acting on the
scalar component of the two-form is
\begin{equation}
\delta_m \hat B_{\alpha\beta} = \delta_m B_{\alpha\beta} =
m_{\alpha\beta},
\end{equation}
where $m_{\alpha\beta}$ is a constant anti-symmetric matrix.
Consistent with the reduction Ansatz the scalars of the Yang-Mills
vectors transform under a ten dimensional Yang-Mills
transformation as
\begin{equation}
\delta_q \hat A^{I}_{\alpha} = \delta_q A^{I}_{\alpha} =
q^{I}_{\alpha},
\end{equation}
where the parameters $q^{I}_{\alpha}$ are constants. To make the
field strength $\hat H = \d \hat B - \tfrac{1}{2}\hat A^I \hat
F^I$ invariant the Kalb-Ramond field must also transform and the
scalar part of the Kalb-Ramond field must transform as
\begin{equation}
\delta_q B_{\alpha\beta} = - q^{I}_{[\alpha} A^{I}_{\beta]} .
\end{equation}
Calculating commutators of these transformations acting on the
scalars we get
\begin{equation}
\begin{array}{l@{\,\quad\,}l}
\left[\delta_{m_1}, \delta_{m_2} \right] = 0 ,& \left[\delta_{q}, \delta_{m} \right] = 0  ,\\
\left[\delta_{\Lambda_1}, \delta_{\Lambda_2} \right] =\delta_{\Lambda_3},& \left[\delta_{\Lambda}, \delta_{q} \right] = -\delta_{q'}  ,\\
\left[\delta_{q_1}, \delta_{q_2} \right] = \delta_{m_3} ,& \left[\delta_{\Lambda}, \delta_{m} \right] =\delta_{m'} ,\\
\Lambda_{3}{}^{\alpha}{}_\beta = \left[\Lambda_1,\Lambda_2 \right]^\alpha{}_\beta , & q'^{I}_{\alpha} = q^{I}_{\gamma}\Lambda^\gamma{}_\alpha  ,\\
m_3 = q_1{}^{I}_{\alpha} q_2{}^{I}_{\beta} -q_2{}^{I}_{\alpha}
q_1{}^{I}_{\beta} ,& m'_{\alpha\beta} =
m_{\alpha\gamma}\Lambda^\gamma{}_\beta
+m_{\gamma\beta}\Lambda^\gamma{}_\alpha .
\end{array}
\end{equation}
We identify generators with these transformations according to
$\Lambda^\alpha{}_{\beta} \leftrightarrow E_{\alpha\beta}$,
$m_{\alpha\beta}\leftrightarrow R_{\alpha\beta}=-R_{\beta\alpha}$
and $q^{I}_{\alpha}\leftrightarrow Y^{I}_{\alpha}$. Note that to
keep $A^{I}{}_{\alpha}$ upper triangular, $\Lambda$ must be an
upper triangular matrix and hence $E_{\alpha\beta}$ only is
nonzero for $\alpha<\beta$. From the given commutators of
transformations we calculate the commutation relations of the
generators and find
\begin{equation}
\begin{array}{l@{\quad}l}
\left[ E_{\alpha\beta}, E_{\gamma\delta} \right]  =
\delta_{\beta\gamma}E_{\alpha\delta} -
\delta_{\alpha\delta}E_{\gamma\beta},&
\left[ R_{\alpha\beta}, R_{\gamma\delta} \right] =0 ,\\
\left[R_{\alpha\beta},Y^{I}_{\gamma} \right] =0 ,&
\left[ Y^{I}_{\alpha}, Y^{J}_{\beta} \right] = 2\delta^{IJ}R_{\alpha\beta}, \\
\left[E_{\alpha\beta},Y^{I}_{\gamma} \right] =
-\delta_{\alpha\gamma}Y^{I}_{\beta} ,& \left[
E_{\alpha\beta},R_{\gamma\delta} \right]  =
-\delta_{\alpha\gamma}R_{\beta\delta} + \delta_{\alpha\delta}
R_{\beta\gamma},
\end{array}
\end{equation}
which is up to a scaling of $R_{\alpha\beta}$ the same as the
algebra given in section \ref{subsecIdent}. Hence the
dimensionally reduced scalar symmetry transformations are
generated by the positive restricted root subalgebra $\mf{n}$ of
$\mf{so}(\d,\d +N)$.

%%%%%%%%%%%%%%%%%%%%%%%%%%%%%%%%%%%%%%%%%%%%%%%%%%%%%%%%%%%%%%%%%%%%%%%%%%%%%%%%%%%%%%%%%
%%%%%%%%%%%%%%%%%%%%%%%%%%%%%%%%%%%%%%%%%%%%%%%%%%%%%%%%%%%%%%%%%%%%%%%%%%%%%%%%%%%%%%%%%
\section{Maximal Scalar Manifolds and
Dualizations}\label{secmaximal} The concept of a maximal scalar
manifold was also discussed in \cite{LuPop}. In $D$ dimensions a
$(D-2)$-form can be dualized to a scalar; when this is done, the
so-called maximal scalar manifold is obtained. For the Heterotic
supergravities this can be done for $D=4$ and $D=3$. In many cases
the global symmetry group is enlarged. If a form is dualized to a
scalar the Lagrangian obtains a topological term, which we will
ignore, and the dilaton coupling vector appears with a different
sign. In four dimensions the dualization of the two-form gives a
term
\begin{equation}
-\tfrac{1}{2} \e^{-\vec A\cdot \vec \Phi} (\partial \chi)^2
\end{equation}
in the Lagrangian and so we can relate to the extra axion $\chi$ a
positive root generator $\mathcal{S}$. Since the vector $\vec A$
has no non-trivial summation rules, $\mathcal{S}$ will commute
with all of them except for the Cartan generators. Hence only the
commutator
\begin{equation}
 [ \vec H \mathcal{S} ] = -\vec A \mathcal{S} = \vec G \mathcal{S}
\end{equation}
is non-zero. One can compare the diagonalization procedure of
\cite{LuPop3} to splitting of this commutator of the
$\mathfrak{so}(\d,\d+N)$ algebra and hence the diagonalization is
also a diagonalization in the Cartan subalgebra of
$\mathfrak{sl}(2;\Real)\oplus\mathfrak{so}(\d,\d+N)$. The
restricted root diagram now consists of two parts, one part is the
same as in section \ref{subsecIdent} with $\d=6$, the other is
that of $SL(2;\Real)/SO(2)\cong SO(2,1)_0/SO(2)$ \footnote{The
index 0 means the component connected to the identity.}. A coset
construction reveals that indeed the global symmetry group is
$SL(2;\Real)\times SO(6,6+N)$ (this is also standard, see e.g.
\cite{MS,LuPop3}). The dualized action can be further reduced to
$D=3$ and in $D=3$ all vector fields can be dualized. One obtains
the following dilaton coupling vectors
\begin{equation}
\begin{split}
& \vec B_{ij},\, -\vec B_i=\vec F_i,\, \vec A_{ij},\,\vec C_{i},\,
 \tfrac{1}{2}\vec G  \, \ 1\leq i,j \leq 7;\\
  &-\vec A_i,\ 1\leq i \leq 6,\  -\vec A_\chi\equiv -\vec A_7,
\end{split}
\end{equation}
where $\vec A_\chi$ results from the vector $\vec A$ in the
dilaton coupling of $\chi$ in four dimensions and we have
\begin{equation}
\begin{array}{ccc}
\vec B_{ij} = -\vec F_i +\vec F_j , & \quad & \vec
A_{ij}=\vec F_i + \vec F_j - \vec G ,\\
\vec C_{i}=\vec F_i -\tfrac{1}{2}\vec G , & \quad & -\vec A_i =
\vec F_i - \vec G .\\
\end{array}
\end{equation}
giving rise to the set of simple restricted roots $\left\{ \vec
B_{i,i+1}, \vec C_1,-\vec A_7 \right\}$. We see that the only
difference with the naively expected $SO(7,7+N)$ symmetry group,
there is an additional restricted root $-\vec A_7$ which only has
a nonzero inner product with $\vec B_{67}$. So we could call it
$\vec B_{78}$ and then we see that the coset structure is the same
as when we would have reduced over 8 dimensions instead of just 7,
and hence the global symmetry group is $SO(8,8+N)$, which is a
known result.

%%%%%%%%%%%%%%%%%%%%%%%%%%%%%%%%%%%%%%%%%%%%%%%%%%%%%%%%%%%%%%%%%%%%%%%%%%%%%%%%%%%%%%%%%
%%%%%%%%%%%%%%%%%%%%%%%%%%%%%%%%%%%%%%%%%%%%%%%%%%%%%%%%%%%%%%%%%%%%%%%%%%%%%%%%%%%%%%%%%

\section{Conclusions}\label{conclusions}

In this paper we outlined a general method for recognizing the
global symmetry groups which appear after dimensional reduction in
extended lower dimensional supergravities. The method can be
broken down into three steps. First one does a circle by circle
dimensional reduction as described in section \ref{sectionmethod}.
As a second step one identifies the dilatonic coupling vectors as
restricted roots and then draws the associated Dynkin diagram and
counts the multiplicities; this fixes the coset. As a third step
one constructs a scalar coset Lagrangian for the coset found in
step two and compares the result to that of the dimensional
reduction.

In this paper we showed that the first and second step uniquely
fix the coset. The complete classification of real forms of simple
Lie algebras turns out to be indispensable and powerful. The third
step is to verify that the scalar part of the Lagrangian obtained
from the reduction coincides with a scalar coset Lagrangian of the
coset found in the second step. The third step can be involved but
contains no difficulties of principles. The method has been
applied to the dimensional reduction of Heterotic supergravity,
where the Lie algebra of $G$ is a non-split real form.

\acknowledgments

We would like to thank Jaap Top for useful discussions and Mees de
Roo for carefully reading the manuscript. The work of DBW is part
of the research programme of the ``Stichting voor Fundamenteel
Onderzoek van de Materie'' (FOM). This work is supported in part
by the European Commission FP6 program MRTN-CT-2004-005104, in
which the Centre for Theoretical Physics in Groningen is
associated to the University of Utrecht.

%%%%%%%%%%%%%%%%%%%%%%%%%%%%%%%%%%%%%%%%%%%%%%%%%%%%%%%%%%%%%%%%%%%%%%%%%%%%%%%%%%%%%%%%%
%%%%%%%%%%%%%%%%%%%%%%%%%%%%%%%%%%%%%%%%%%%%%%%%%%%%%%%%%%%%%%%%%%%%%%%%%%%%%%%%%%%%%%%%%
%%%%%%%%%%%%%                 APPENDICES                                  %%%%%%%%%%%%%%%
%%%%%%%%%%%%%%%%%%%%%%%%%%%%%%%%%%%%%%%%%%%%%%%%%%%%%%%%%%%%%%%%%%%%%%%%%%%%%%%%%%%%%%%%%
%%%%%%%%%%%%%%%%%%%%%%%%%%%%%%%%%%%%%%%%%%%%%%%%%%%%%%%%%%%%%%%%%%%%%%%%%%%%%%%%%%%%%%%%%

\begin{appendices}

%%%%%%%%%%%%%%%%%%%%%%%%%%%%%%%%%%%%%%%%%%%%%%%%%%%%%%%%%%%%%%%%%%%%%%%%%%%%%%%%%%%%%%%%%

\section{Real Forms of Lie algebras: a Quick
Reference}\label{appendixLie1}

Every semi-simple real Lie algebra $\mf{g}$ admits a Cartan
involution $\theta$ which is an involutive automorphism, such that
the bilinear form $B_\theta(x,y)= - B(x,\theta y)$ is positive
definite and $B$ denotes the Cartan-Killing form. The $+1(-1)$
eigenspace is denoted $\mf{k}(\mf{p})$ and we have the Cartan
decomposition $\mf{g}=\mf{k}\oplus\mf{p}$ and schematically the
commutation relations are as in eqn.(\ref{commutatiecompact}).
With respect to the inner product $B_\theta$ we have $(\ad
x)^\dagger = -\ad \theta x$ and thus a maximal abelian subalgebra
$\mf{a}\subset \mf{p}$ induces a decomposition of $\mf{g}$ into
eigenspaces w.r.t. $\mf{a}$. This is a restricted root
decomposition analogous to the usual root decomposition and we
have the ortogonal decomposition
\begin{equation}
\begin{split}
\mf{g}&=\mf{g}_0\oplus\bigoplus_{\lambda\in\Sigma}\mf{g}_\lambda,
\quad  \mf{g}_0 = \mf{a}\oplus Z_\mf{k}(\mf{a}) ,\\
[Hx_\lambda]&=\lambda(H)x_\lambda, \forall
H\in\mf{a},x_\lambda\in\mf{g}_\lambda,
\end{split}
\end{equation}
where $\Sigma\subset\mf{a}^*$ denotes the set of restricted roots
and $Z_\mf{k}(\mf{a})$ is the centralizer of $\mf{a}$ in $\mf{k}$.
The Iwasawa decomposition of the Lie algebra $\mf{g}$ is
\begin{equation}
\mathfrak{g} = \mathfrak{k} \oplus \mathfrak{a} \oplus
\mathfrak{n}, \ \ \mathfrak{n} = \bigoplus_{\lambda \in \Sigma^+}
\mathfrak{g}_\lambda .
\end{equation}

Let $\mf{t}$ be a maximal torus in $Z_\mf{k}(\mf{a})$.
Complexifying the Lie algebra to $\mf{g}^\mathbbm{C}$ a Cartan
subalgebra $\mf{h}^\mathbbm{C}$ can be found by extending
$\mf{a}^\mathbbm{C}$ with the complexified maximal torus
$\mf{t}^\mathbbm{C}$. The set of roots is denoted $\Delta$ and the
roots are real on $\mf{a}\oplus i\mf{t}$. The action of $\theta$
is extended by linearity and we see that $-\theta$ acts as complex
conjugation on the real Cartan subalgebra $\mf{h}=\mf{a}\oplus i
\mf{t}$. The action of $\theta$ on a root $\alpha$ is defined by
$(\theta\alpha)(h)= \alpha(\theta h), \ h\in\mf{h}$. A root
$\alpha$ is called real if $\alpha|_{i\mathfrak{t}}\equiv 0$, and
imaginary of $\alpha|_{\mathfrak{a}}=0$ and complex if it is
neither real nor complex. One can find an ordering of the roots in
which $\mf{a}$ is taken before $i\mf{t}$ implying that $\theta$
permutes the simple roots. A Satake diagram is a diagram in which
the imaginary simple roots are colored, while the real and complex
are not but the two-element orbits of $\theta$ is denoted by
arrows. The projection of a root $\alpha$ to its restriction to
$\mf{a}$, denoted $\bar \alpha$, is easily seen to be $ \bar
\alpha = \tfrac{1}{2}(\alpha - \theta\alpha) $. It is easy to see
that $\bar \alpha (i\mf{t})=0$ and $\bar\alpha\in\Sigma$. If
$\lambda$ is a restricted root, then we define
$\Delta_\lambda=\left\{ \alpha \in \Delta | \bar \alpha
=\lambda\right\}$. The number of roots in this subspace is called
the multiplicity of the restricted root $\lambda$:
$m_{\lambda}=\textrm{Card}\Delta_\lambda$. Since for restricted
roots the root system can be non-reduced (i.e. if $\lambda\in
\Sigma$, also $2\lambda$ can be in $\Sigma$) also $m_{2\lambda}$
can be non-zero.

A simple restricted root is a positive restricted root which can
not be written as the sum of two positive restricted roots. The
set of simple restricted roots
$\left\{\lambda_1,\ldots,\lambda_l\right\}$ contains
$l=\dim\mf{a}$ elements and gives rise to a restricted root Dynkin
diagram. It is a theorem that if the restricted root system (i.e.
the restricted root Dynkin diagram) and the multiplicities
$m_{\lambda_i}$ and $m_{2\lambda_i}$ of the simple restricted
roots $\lambda_i$ are known, the real form of the simple Lie
algebra is uniquely determined. This enables us to list all of
them and since the compact form always exists, we only list all
the non-compact real forms in appendix \ref{tables}. In these
tables the simple roots $\alpha_i,\,1\leq i \leq r =\dim_\Real
\mf{h}$ are related to the simple restricted roots $\lambda_i$ by
$\lambda_i = \bar\alpha_i$. The table contains information that
can be found in \cite{helgason}. The cosets $G/K$ with $K$
maximally compact are also put in the table. The Satake diagrams
are there to show which simple roots of the original Dynkin
diagram survive the projection to the simple restricted roots.

%%%%%%%%%%%%%%%%%%%%%%%%%%%%%%%%%%%%%%%%%%%%%%%%%%%%%%%%%%%%%%%%%%%%%%%%%%%%%%%%%%%%%%%%%
%%%%%%%%%%%%%%%%%%%%%%%%%%%%%%%%%%%%%%%%%%%%%%%%%%%%%%%%%%%%%%%%%%%%%%%%%%%%%%%%%%%%%%%%%
\section{Some details of the Lie algebra $\mathfrak{so}(\d, \d +
N )$.} \label{appendixLie2}

In this section we use the same notation as in appendix
\ref{appendixLie1}. With the use of the $\eta$-matrix
\begin{equation}
\eta = \begin{pmatrix} 0& \mathbbm{1}_{\d\times \d} & 0 \\
\mathbbm{1}_{\d\times \d}& 0 & 0\\ 0&0&\mathbbm{1}_{N\times N} \\
\end{pmatrix}
\end{equation}
it is easy to work out the constraint $X^T \eta + \eta X=0$ for a
$(2\d + N)\times (2\d +N)$-matrix $X$. The Cartan involution
$\theta$ will then be defined by its action in this vector
representation $\mathfrak{so}(\d,\d+N)_V$ by
\begin{equation}
 \theta (X) \mapsto -X^\dagger=-X^T ; X \in
 \mathfrak{so}(\d,\d+N)_V,
\end{equation}
and since the vector representation is faithful, this is indeed an
automorphism of the Lie algebra. The Cartan decomposition
$\mf{g}=\mf{k}\oplus\mf{p}$ into the eigenspaces of $\theta$ is
easily found. A maximal abelian subspace $\mathfrak{a}$ of
$\mf{p}$ where $\theta$ is $-\id$ can be found to be spanned by
the diagonal matrices of the form
\begin{equation}
A = \begin{pmatrix} a & 0 & 0 \\ 0 & -a & 0 \\ 0 & 0 & 0 \\
\end{pmatrix},
\end{equation}
where $a$ is $\d\times\d$ diagonal matrix. We can set up an
isomorphism between $\Real^\d$ and $\mathfrak{a}$ by the
isomorphism (which is a vector-space isomorphism) $\phi: \Real^\d
\rightarrow \mathfrak{a}$ defined by
\begin{equation}
\phi: \vec a = (a_1,\ldots,a_\d) \mapsto
diag(a_1,\ldots,a_\d,-a_1,\ldots,-a_\d,\underbrace{0,\ldots,0}_{N
\textrm{ zeros }} ),
\end{equation}
Now define the following basis in the dual of $\Real^\d$
\begin{equation}
\tilde \lambda_i : \vec a = (a_1,\ldots,a_\d) \mapsto a_i \in
\Real .
\end{equation}
The restricted roots of $\mf{so}(\d,\d+N)$ with respect to
$\mathfrak{a}$ can be calculated to be
\begin{equation}
0, \ \pm \lambda_i\equiv \pm\phi\circ\tilde \lambda_i
\circ\phi^{-1}, \ \pm (\lambda_i-\lambda_j) \ i \neq j, \
\pm(\lambda_i+\lambda_j)\ i \neq j,
\end{equation}
and denoting the restricted root spaces by $\mathfrak{g}_\lambda$
for a restricted root $\lambda$, we have
dim$\mathfrak{g}_{\pm\lambda_i}=N$, dim$\mathfrak{g}_{\pm(
\lambda_i \pm\lambda_j)}=1$. We can see that $E_{ij}\in
\mathfrak{g}_{\lambda_i-\lambda_j}$, $R_{ij} \in
\mathfrak{g}_{-\lambda_i-\lambda_j}$ and $Y_{iI}\in
\mathfrak{g}_{-\lambda_i}$ where
\begin{equation}
E_{ij}  = \begin{pmatrix}
                     \phantom{-}e_{ij}&\phantom{-}0&\phantom{-}0 \\
                     \phantom{-}0& -e_{ji} & \phantom{-}0\\
                     \phantom{-}0&\phantom{-}0&\phantom{-}0\\
                     \end{pmatrix}, \,
R_{ij}  = \begin{pmatrix}
                     \phantom{-}0&\phantom{-}0&\phantom{-}0\\
                     \phantom{-}\beta_{ji}&\phantom{-}0&\phantom{-}0\\
                     \phantom{-}0&\phantom{-}0&\phantom{-}0\\
                     \end{pmatrix}, \,
Y_{iI}  = \begin{pmatrix}
                     \phantom{-}0&\phantom{-}0&\phantom{-}0\\
                     \phantom{-}0&\phantom{-}0&\phantom{-}\gamma_{iI}\\
                     -\gamma_{iI}{}^T&\phantom{-}0&\phantom{-}0&\\
                     \end{pmatrix},
\end{equation}
with
\begin{subequations}
\begin{align}
(e_{ij})_{kl} &= \delta_{ik}\delta_{jl} , \ 1\leq i,j,k,l \leq \d ,\\
\beta_{ij} &=e_{ij} - e_{ji}, \ 1\leq i<j \leq \d ,\\
E_{ij} &= e_{ij}, \ 1\leq i<j \leq \d ,\\
(\gamma_{iI})_{kK} &= \delta_{ik}\delta_{IK} ,\ 1\leq i,k \leq
\d,\, 1\leq I,K\leq N.
\end{align}
\end{subequations}
Having chosen a sense of positivity and calling $\Sigma^+$ the set
of positive roots, the Iwasawa decomposition is then
\begin{equation}
\mathfrak{g} = \mathfrak{k}\oplus \mathfrak{a} \oplus \mathfrak{n}
, \quad \mathfrak{n}=\bigoplus_{\lambda \in \Sigma^+}
\mathfrak{g}_\lambda .
\end{equation}
We see that we can chose $\mathfrak{n}$ to be the span of the
union of the sets $\left\{E_{ij}: 1\leq i<j\leq \d \right\}$,
$\left\{ R_{ij} = -R_{ji} : 1 \leq i,j\leq\d\right\}$ and $\left\{
Y_{iI} : 1\leq i \leq \d, 1 \leq I \leq N \right\}$. Though some
minus signs may favor the name negative root part, this is just
merely a matter of choice. A quick calculation reveals:
\begin{equation}
\begin{array}{l@{\quad}l}
\left[ E_{i j} Y_{k K} \right] = - \delta_{ik} Y_{j K } ,& \left[Y_{i I} Y_{j J} \right] = \delta_{IJ} R_{ij} ,\\
\left[ E_{ij} E_{kl} \right] = \delta_{jk} E_{il} -\delta_{il} E_{kj} ,& \left[E_{ij} R_{kl} \right]= - \delta_{ik} R_{jl} + \delta_{il} R_{jk} ,\\
\left[ E_{i j} Y_{k K} \right] = - \delta_{ik} Y_{j K } ,& \left[Y_{i I} Y_{j J} \right] = \delta_{IJ} R_{ij} ,\\
\left[Y_{iI}R_{kl}\right] =0 ,& \left[R_{ij}R_{kl}\right]=0 .
\end{array}
\end{equation}

%%%%%%%%%%%%%%%%%%%%%%%%%%%%%%%%%%%%%%%%%%%%%%%%%%%%%%%%%%%%%%%%%%%%%%%%%%%%%%%%%%%%%%%%%
\section{Baker-Campbell-Hausdorff and tricks}\label{appendixBCH}
Useful formulae are
\begin{subequations}
\begin{align}
\e^X Y\e^{-X} &=\textrm{Ad}(\e^X)Y = \e^{\textrm{ad}X}Y =
Y+[X,Y]+\tfrac{1}{2}[X,[X,Y]]+\ldots ,\\
\left(\d \e^X\right)  e^{-X} &= \d X + \tfrac{1}{2}[ X , \d X ] +
\tfrac{1}{6}[X,[X,\d X ] ]+\ldots ,
\end{align}
\end{subequations}
for explanations see \cite{hall}. In deriving the formulas
\ref{trick1}-\ref{trick3} it is handy if one uses that in the
fundamental representation of the $E_{ij}$ we have
$(\mathcal{V}_2)^{-1}_{ij} = \Gamma_{ij}$, so we get $\d
(\mathcal{V}_2)_{jk} \Gamma_{km} = \d A_{jk} \Gamma_{km} =
F_{jm}$. The fundamental representation is faithful and hence it
holds for all representations, hence for the Lie algebra. It is
handy to note that $E_{ij}$ acts on the $Y_{iI}$ as a linear
transformation in the vector space spanned by the $Y_i$ - omitting
the $I$-index since it is in this case a spectator:
\begin{equation}
\ad E_{ij} (Y_{n}) = -\delta_{in}Y_{j}= \sum_m(\ad
E_{ij})_{nm}Y_{m} \ \Rightarrow \ (\ad E_{ij})_{mn} =
-\delta_{im}\delta_{jn}.
\end{equation}
The representing matrix of $\ad E_{ij}$ squares to zero giving
$\ad E_{ij}\circ \ad E_{ij} (Y_{n}) =0$. One can easily proceed
via
\begin{equation}
\e^{A_{ij}E_{ij}} Y_{n} \e^{-A_{ij}E_{ij}} =
\textrm{Ad}(\e^{A_{ij}E_{ij}}) ( Y_{n} ) = \e^{\ad
A_{ij}E_{ij}}(Y_{n}) = (\mathbbm{1} + \ad A_{ij}E_{ij})_{nm}Y_{m}.
\end{equation}
The generators $R_{mn}$ are in the tensor-product representation
of this representation under the action of $\ad E_{ij}$ and hence
one finds:
\begin{equation}
\mathcal{V}_2 R_{mn} \mathcal{V}_2{}^{-1} =
\Gamma_{mp}\Gamma_{nq}R_{pq} .
\end{equation}

%%%%%%%%%%%%%%%%%%%%%%%%%%%%%%%%%%%%%%%%%%%%%%%%%%%%%%%%%%%%%%%%%%%%%%%%%%%%%%%%%%%%%%%%%
%%%%%%%%%%%%%%%%%%%%%%%%%%%%%%%%%%%%%%%%%%%%%%%%%%%%%%%%%%%%%%%%%%%%%%%%%%%%%%%%%%%%%%%%%
\section{Tables}\label{tables}

\begin{longtable}{cc@{\quad}l}
\caption[Table 1]{Satake diagrams and restricted root diagrams and associated cosets $G/K$.}\\
\hline Satake Diagram & Restricted Root Diagram & Type \\
\hline \endfirsthead
\caption[Table 1]{continued}\\
\hline Satake Diagram & Restricted Root Diagram & Type \\
\hline \endhead \hline \multicolumn{3}{r}{Continued on next page}
\endfoot
\hline\endlastfoot
\begin{pspicture}(0,0)(5,2)
\pscircle(0.5,1){0.1}%1
\qline(0.7,1)(1.2,1)\pscircle(1.4,1){0.1}%2
\qline(1.6,1)(2.1,1)\psline[linestyle=dotted,dotsep=0.1,linewidth=0.05](2.2,1)(2.7,1)%
\qline(2.8,1)(3.4,1)\pscircle(3.6,1){0,1}%3
\qline(3.8,1)(4.3,1)\pscircle(4.5,1){0.1}%4
\put(0.4,1.2){\f{\alpha_1}}%1
\put(1.3,1.2){\f{\alpha_2}}%2
\put(3.5,1.2){\f{\alpha_{r-1}}}%3
\put(4.4,1.2){\f{\alpha_r}}%4
\end{pspicture}
&
\begin{pspicture}(0,0)(5,2)
\pscircle(0.5,1){0.1}%1
\qline(0.7,1)(1.2,1)\pscircle(1.4,1){0.1}%2
\qline(1.6,1)(2.1,1)\psline[linestyle=dotted,dotsep=0.1,linewidth=0.05](2.2,1)(2.7,1)%
\qline(2.8,1)(3.4,1)\pscircle(3.6,1){0,1}%3
\qline(3.8,1)(4.3,1)\pscircle(4.5,1){0.1}%4
\put(0.4,1.2){\f{\lambda_1}}%1
\put(1.3,1.2){\f{\lambda_2}}%2
\put(3.5,1.2){\f{\lambda_{r-1}}}%3
\put(4.4,1.2){\f{\lambda_r}}%4
\end{pspicture}
&
\begin{pspicture}(0,0)(3,2)
\put(0,1.4){\ft{\textbf{A\textit{I}}:$SL(n;\Real)/SO(n)$}}\put(0,0.7){\f{l=r=n-1}}%
\end{pspicture} \\
\begin{pspicture}(0,0)(6,2)
\pscircle*(0.5,1){0.1}%1
\qline(0.7,1)(1.2,1)\pscircle(1.4,1){0.1}%2
\qline(1.6,1)(2.1,1)\pscircle*(2.3,1){0.1}%3
\qline(2.5,1)(3.0,1)\psline[linestyle=dotted,dotsep=0.1,linewidth=0.05](3.1,1)(3.6,1)%
\qline(3.7,1)(4.2,1)\pscircle(4.4,1){0.1}%2l
\qline(4.6,1)(5.1,1)\pscircle*(5.3,1){0.1}%r
\put(0.4,1.2){\f{\alpha_1}}%1
\put(1.3,1.2){\f{\alpha_2}}%2
\put(4.3,1.2){\f{\alpha_{2l}}}%2l
\put(5.2,1.2){\f{\alpha_r}}% r
\end{pspicture}
&
\begin{pspicture}(0,0)(5,2)
\pscircle(0.5,1){0.1}%1
\qline(0.7,1)(1.2,1)\pscircle(1.4,1){0.1}%2
\qline(1.6,1)(2.1,1)\psline[linestyle=dotted,dotsep=0.1,linewidth=0.05](2.2,1)(2.7,1)%
\qline(2.8,1)(3.4,1)\pscircle(3.6,1){0,1}%3
\qline(3.8,1)(4.3,1)\pscircle(4.5,1){0.1}%4
\put(0.4,1.2){\f{\lambda_2}}%1
\put(1.3,1.2){\f{\lambda_4}}%2
\put(3.5,1.2){\f{\lambda_{2l-2}}}%3
\put(4.4,1.2){\f{\lambda_{2l}}}%4
\end{pspicture}
&
\begin{pspicture}(0,0)(3,2)
\put(0,1.4){\ft{\textbf{A\textit{II}}:$SU^*(2n)/Sp(n)$}}\put(0,0.7){\f{l=2r-l=n-1}}%
\end{pspicture} \\
\begin{pspicture}(0,0)(5.2,4)
\pscircle(0.5,3.3){0.1}%1
\qline(0.7,3.3)(1.2,3.3)\pscircle(1.4,3.3){0.1}%2
\qline(1.6,3.3)(2.1,3.3)\psline[linestyle=dotted,dotsep=0.1,linewidth=0.05](2.2,3.3)(2.7,3.3)%
\qline(2.8,3.3)(3.4,3.3)\pscircle(3.6,3.3){0,1}%3
\qline(3.8,3.3)(4.3,3.3)\pscircle*(4.5,3.3){0.1}%4
\qline(4.5,3.1)(4.5,2.6)\pscircle*(4.5,2.4){0.1}%5
\psline[linestyle=dotted,dotsep=0.1,linewidth=0.05](4.5,2.2)(4.5,1.7)\pscircle*(4.5,1.5){0.1}%
\qline(4.5,1.3)(4.5,0.8)\pscircle*(4.5,0.6){0.1}%
\pscircle(0.5,0.6){0.1}%1'
\qline(0.7,0.6)(1.2,0.6)\pscircle(1.4,0.6){0.1}%2'
\qline(1.6,0.6)(2.1,0.6)\psline[linestyle=dotted,dotsep=0.1,linewidth=0.05](2.2,0.6)(2.7,0.6)%
\qline(2.8,0.6)(3.4,0.6)\pscircle(3.6,0.6){0,1}%3'
\qline(3.8,0.6)(4.3,0.6)%
\put(0.4,3.5){\f{\alpha_1}}%1
\put(1.3,3.5){\f{\alpha_2}}%2
\put(3.5,3.5){\f{\alpha_{l}}}%3
\psline[linearc=3]{<->}(0.4,3.15)(0.2,1.95)(0.40,0.75)%
\psline[linearc=3]{<->}(1.30,3.15)(1.1,1.95)(1.3,0.75)%
\psline[linearc=3]{<->}(3.5,3.15)(3.3,1.95)(3.5,0.75)%
\end{pspicture}
&
\begin{pspicture}(0,-1)(5,2)
\pscircle(0.5,1){0.1}%1
\qline(0.7,1)(1.2,1)\pscircle(1.4,1){0.1}%2
\qline(1.6,1)(2.1,1)\psline[linestyle=dotted,dotsep=0.1,linewidth=0.05](2.2,1)(2.7,1)%
\qline(2.8,1)(3.4,1)\pscircle(3.6,1){0,1}%3
\psline[doubleline=true,doublesep=0.03,arrows=->,arrowlength=0.6,arrowinset=0.55](3.8,1)(4.3,1)\pscircle(4.5,1){0.1}%4
\put(0.4,1.2){\f{\lambda_1}}%1
\put(1.3,1.2){\f{\lambda_2}}%2
\put(3.5,1.2){\f{\lambda_{r-1}}}%3
\put(4.4,1.2){\f{\lambda_r}}%4
\end{pspicture}
&
\begin{pspicture}(0,0)(3,4)
\put(0,3.5){\ft{\textbf{A\textit{III}}:$SU(p,q)/S(U_p\times
U_q)$}}%
\put(0,3.0){\f{l=min(p,q), \, r=p+q-1}}%
\put(0,2){\ft{If $2\leq l \leq r/2$}}%
\end{pspicture}
\\
\begin{pspicture}(0,1.7)(5.2,4)
\pscircle(0.5,3.3){0.1}%1
\qline(0.7,3.3)(1.2,3.3)\pscircle(1.4,3.3){0.1}%2
\qline(1.6,3.3)(2.1,3.3)\psline[linestyle=dotted,dotsep=0.1,linewidth=0.05](2.2,3.3)(2.7,3.3)%
\qline(2.8,3.3)(3.4,3.3)\pscircle(3.6,3.3){0,1}%3
\qline(3.74,3.16)(4.09,2.81)\pscircle(4.23,2.67){0.1}% 44
\qline(4.09,2.53)(3.74,2.18)%
\pscircle(0.5,2.04){0.1}%1'
\qline(0.7,2.04)(1.2,2.04)\pscircle(1.4,2.04){0.1}%2'
\qline(1.6,2.04)(2.1,2.04)\psline[linestyle=dotted,dotsep=0.1,linewidth=0.05](2.2,2.04)(2.7,2.04)%
\qline(2.8,2.04)(3.4,2.04)\pscircle(3.6,2.04){0,1}%3'
\put(0.4,3.5){\f{\alpha_1}}%1
\put(1.3,3.5){\f{\alpha_2}}%2
\put(4.13,2.87){\f{\alpha_{l}}}%3
\psline[linearc=1]{<->}(0.4,3.15)(0.3,2.65)(0.4,2.14)%
\psline[linearc=1]{<->}(1.3,3.15)(1.2,2.65)(1.3,2.14)%
\psline[linearc=1]{<->}(3.5,3.15)(3.4,2.65)(3.5,2.14)%
\end{pspicture}
&
\begin{pspicture}(0,0)(5,2)
\pscircle(0.5,1){0.1}%1
\qline(0.7,1)(1.2,1)\pscircle(1.4,1){0.1}%2
\qline(1.6,1)(2.1,1)\psline[linestyle=dotted,dotsep=0.1,linewidth=0.05](2.2,1)(2.7,1)%
\qline(2.8,1)(3.4,1)\pscircle(3.6,1){0,1}%3
\psline[doubleline=true,doublesep=0.03,arrows=<-,arrowlength=0.6,arrowinset=0.55](3.8,1)(4.3,1)\pscircle(4.5,1){0.1}%4
\put(0.4,1.2){\f{\lambda_1}}%1
\put(1.3,1.2){\f{\lambda_2}}%2
\put(3.5,1.2){\f{\lambda_{r-1}}}%3
\put(4.4,1.2){\f{\lambda_r}}%4
\end{pspicture}
&
\begin{pspicture}(0,0)(3,2)
\put(0,1){\ft{If $r=2l-1$}}
\end{pspicture}
\\
\begin{pspicture}(0,0)(5,2)
\pscircle(0.5,1){0.1}%1
\qline(0.7,1)(1.2,1)\pscircle*(1.4,1){0.1}%2
\qline(1.6,1)(2.1,1)\psline[linestyle=dotted,dotsep=0.1,linewidth=0.05](2.2,1)(2.7,1)%
\qline(2.8,1)(3.4,1)\pscircle*(3.6,1){0,1}%3
\qline(3.8,1)(4.3,1)\pscircle(4.5,1){0.1}%4
\put(0.4,1.2){\f{\alpha_1}}%1
\put(4.4,1.2){\f{\alpha_r}}%4
\psline[linearc=7]{<->}(0.6,0.9)(2.5,0.7)(4.4,0.9)%
\end{pspicture}
&
\begin{pspicture}(-1,-1)(1,1)
\pscircle(0,0){0.1}\put(-0.1,0.2){\f{\lambda_1}}%
\end{pspicture}
&
\begin{pspicture}(0,0)(3,2)
\put(0,1.4){\ft{\textbf{A\textit{IV}}:$SU(n,1)/SU(n)$}}\put(0,0.7){\f{l=r+1=n}}%
\end{pspicture}
\\
\begin{pspicture}(0.4,0)(6.5,2)
\pscircle(0.5,1){0.1}%1
\qline(0.7,1)(1.2,1)\psline[linestyle=dotted,dotsep=0.1,linewidth=0.05](1.3,1)(1.8,1)%
\qline(1.9,1)(2.4,1)\pscircle(2.6,1){0.1}%
\qline(2.8,1)(3.5,1)\pscircle*(3.7,1){0.1}\qline(3.9,1)(4.4,1)%
\psline[linestyle=dotted,dotsep=0.1,linewidth=0.05](4.5,1)(5,1)%
\pscircle*(5.2,1){0.1}\psline[doubleline=true,doublesep=0.03,arrows=->,arrowlength=0.6,arrowinset=0.55](5.4,1)(5.9,1)%
\pscircle*(6.1,1){0.1}%
\put(0.4,1.2){\f{\alpha_1}}\put(2.5,1.2){\f{\alpha_l}}%
\end{pspicture}
&
\begin{pspicture}(0,0)(5,2)
\pscircle(0.5,1){0.1}%1
\qline(0.7,1)(1.2,1)\pscircle(1.4,1){0.1}%2
\qline(1.6,1)(2.1,1)\psline[linestyle=dotted,dotsep=0.1,linewidth=0.05](2.2,1)(2.7,1)%
\qline(2.8,1)(3.4,1)\pscircle(3.6,1){0,1}%3
\psline[doubleline=true,doublesep=0.03,arrows=->,arrowlength=0.6,arrowinset=0.55](3.8,1)(4.3,1)\pscircle(4.5,1){0.1}%4
\put(0.4,1.2){\f{\lambda_1}}%1
\put(4.4,1.2){\f{\lambda_l}}%4
\end{pspicture}
&
\begin{pspicture}(0,0)(3,2)
\put(0,1.4){\ft{\textbf{B\textit{I}}:$\frac{SO(p,q)}{SO(p)\times SO(q)}; \, p+q$ odd}}\put(0,0.7){\f{2\leq l=min(p,q)\leq r}}%
\end{pspicture}
 \\
\begin{pspicture}(0,0)(5,2)
\pscircle(0.5,1){0.1}%1
\qline(0.7,1)(1.2,1)\pscircle*(1.4,1){0.1}%2
\qline(1.6,1)(2.1,1)\psline[linestyle=dotted,dotsep=0.1,linewidth=0.05](2.2,1)(2.7,1)%
\qline(2.8,1)(3.4,1)\pscircle*(3.6,1){0,1}%3
\psline[doubleline=true,doublesep=0.03,arrows=->,arrowlength=0.6,arrowinset=0.55](3.8,1)(4.3,1)\pscircle*(4.5,1){0.1}%4
\put(0.4,1.2){\f{\alpha_1}}%1
\end{pspicture}
&
\begin{pspicture}(-1,-1)(1,1)
\pscircle(0,0){0.1}\put(-0.1,0.2){\f{\lambda_1}}%
\end{pspicture}
&
\begin{pspicture}(0,0)(3,2)
\put(0,1.4){\ft{\textbf{B\textit{II}}:$SO(2p,1)/SO(2p)$}}\put(0,0.7){\f{l=1, \, r=p}}%
\end{pspicture}
\\
\begin{pspicture}(0,0)(5,2)
\pscircle(0.5,1){0.1}%1
\qline(0.7,1)(1.2,1)\pscircle(1.4,1){0.1}%2
\qline(1.6,1)(2.1,1)\psline[linestyle=dotted,dotsep=0.1,linewidth=0.05](2.2,1)(2.7,1)%
\qline(2.8,1)(3.4,1)\pscircle(3.6,1){0,1}%3
\psline[doubleline=true,doublesep=0.03,arrows=<-,arrowlength=0.6,arrowinset=0.55](3.8,1)(4.3,1)\pscircle(4.5,1){0.1}%4
\put(0.4,1.2){\f{\alpha_1}}%1
\put(1.3,1.2){\f{\alpha_2}}%2
\put(3.5,1.2){\f{\alpha_{r-1}}}%3
\put(4.4,1.2){\f{\alpha_r}}%4
\end{pspicture}
&
\begin{pspicture}(0,0)(5,2)
\pscircle(0.5,1){0.1}%1
\qline(0.7,1)(1.2,1)\pscircle(1.4,1){0.1}%2
\qline(1.6,1)(2.1,1)\psline[linestyle=dotted,dotsep=0.1,linewidth=0.05](2.2,1)(2.7,1)%
\qline(2.8,1)(3.4,1)\pscircle(3.6,1){0,1}%3
\psline[doubleline=true,doublesep=0.03,arrows=<-,arrowlength=0.6,arrowinset=0.55](3.8,1)(4.3,1)\pscircle(4.5,1){0.1}%4
\put(0.4,1.2){\f{\lambda_1}}%1
\put(1.3,1.2){\f{\lambda_2}}%2
\put(3.5,1.2){\f{\lambda_{r-1}}}%3
\put(4.4,1.2){\f{\lambda_r}}%4
\end{pspicture}
&
\begin{pspicture}(0,0)(3,2)
\put(0,1.4){\ft{\textbf{C\textit{I}}:$Sp(n,\Real)/U(n)$}}\put(0,0.7){\f{l=r=n}}%
\end{pspicture}
\\
\begin{pspicture}(0.2,0)(7.1,2)
\pscircle*(0.5,1){0.1}%1
\qline(0.7,1)(1.2,1)\pscircle(1.4,1){0.1}%2
\qline(1.6,1)(2.1,1)\pscircle*(2.3,1){0.1}%3
\qline(2.5,1)(3.0,1)\psline[linestyle=dotted,dotsep=0.1,linewidth=0.05](3.1,1)(3.6,1)%
\pscircle(3.8,1){0.1}%2l
\qline(4,1)(4.5,1)\pscircle*(4.7,1){0.1}%2l+1
\qline(4.9,1)(5.3,1)\psline[linestyle=dotted,dotsep=0.1,linewidth=0.05](5.4,1)(5.8,1)%
\pscircle*(6.0,1){0.1}\psline[doubleline=true,doublesep=0.03,arrows=<-,arrowlength=0.6,arrowinset=0.55](6.2,1)(6.7,1)%
\pscircle*(6.9,1){0.1}%
\put(0.4,1.2){\f{\alpha_1}}%1
\put(1.3,1.2){\f{\alpha_2}}%2
\put(3.7,1.2){\f{\alpha_{2l}}}%2l
\end{pspicture}
&
\begin{pspicture}(0,0)(5,2)
\pscircle(0.5,1){0.1}%1
\qline(0.7,1)(1.2,1)\pscircle(1.4,1){0.1}%2
\qline(1.6,1)(2.1,1)\psline[linestyle=dotted,dotsep=0.1,linewidth=0.05](2.2,1)(2.7,1)%
\qline(2.8,1)(3.4,1)\pscircle(3.6,1){0,1}%3
\psline[doubleline=true,doublesep=0.03,arrows=->,arrowlength=0.6,arrowinset=0.55](3.8,1)(4.3,1)\pscircle(4.5,1){0.1}%4
\put(0.4,1.2){\f{\lambda_2}}%1
\put(1.3,1.2){\f{\lambda_4}}%2
\put(4.4,1.2){\f{\lambda_{2l}}}%4
\end{pspicture}
&
\begin{pspicture}(0,0)(3,2.65)
\put(0,2.1){\ft{\textbf{C\textit{II}}:$\frac{Sp(p,q)}{S(p)\times Sp(q)}$}}%
\put(0,1.5){\f{l=min(p,q)}}%
\put(0,1.0){\ft{If $1\leq l \leq \tfrac{1}{2}(r-l)$}}%
\end{pspicture}
\\
\begin{pspicture}(0,0)(6,2)
\pscircle*(0.5,1){0.1}%1
\qline(0.7,1)(1.2,1)\pscircle(1.4,1){0.1}%2
\qline(1.6,1)(2.1,1)\pscircle*(2.3,1){0.1}%3
\qline(2.5,1)(3.0,1)\psline[linestyle=dotted,dotsep=0.1,linewidth=0.05](3.1,1)(3.6,1)%
\qline(3.7,1)(4.2,1)\pscircle(4.4,1){0.1}%4
\qline(4.6,1)(5.1,1)\pscircle*(5.3,1){0.1}%5
\psline[doubleline=true,doublesep=0.03,arrows=<-,arrowlength=0.6,arrowinset=0.55](5.5,1)(6.0,1)%
\pscircle(6.2,1){0.1}%
\put(0.4,1.2){\f{\alpha_1}}%1
\put(1.3,1.2){\f{\alpha_2}}%2
\put(6.1,1.2){\f{\alpha_{2l}}}%2l
\end{pspicture}
&
\begin{pspicture}(0,0)(5,2)
\pscircle(0.5,1){0.1}%1
\qline(0.7,1)(1.2,1)\pscircle(1.4,1){0.1}%2
\qline(1.6,1)(2.1,1)\psline[linestyle=dotted,dotsep=0.1,linewidth=0.05](2.2,1)(2.7,1)%
\qline(2.8,1)(3.4,1)\pscircle(3.6,1){0,1}%3
\psline[doubleline=true,doublesep=0.03,arrows=<-,arrowlength=0.6,arrowinset=0.55](3.8,1)(4.3,1)\pscircle(4.5,1){0.1}%4
\put(0.4,1.2){\f{\lambda_2}}%1
\put(1.3,1.2){\f{\lambda_4}}%2
\put(4.4,1.2){\f{\lambda_{2l}}}%4
\end{pspicture}
&
\begin{pspicture}(0,0)(3,2)
\put(0,1){\ft{If $2\leq l = r/2$}}
\end{pspicture}
\\
\begin{pspicture}(0,0)(6,2)
\pscircle(0.5,1){0.1}%1
\qline(0.7,1)(1.2,1)\psline[linestyle=dotted,dotsep=0.1,linewidth=0.05](1.3,1)(1.8,1)%
\qline(1.9,1)(2.4,1)\pscircle(2.6,1){0.1}%
\qline(2.8,1)(3.5,1)\pscircle*(3.7,1){0.1}\qline(3.9,1)(4.4,1)%
\psline[linestyle=dotted,dotsep=0.1,linewidth=0.05](4.5,1)(5,1)%
\pscircle*(5.2,1){0.1}\qline(5.34,1.14)(5.69,1.39)\qline(5.34,0.86)(5.69,0.51)%
\pscircle*(5.83,1.53){0.1}\pscircle*(5.83,0.37){0.1}%
\put(0.4,1.2){\f{\alpha_1}}\put(2.5,1.2){\f{\alpha_l}}%
\end{pspicture}
&
\begin{pspicture}(0,0)(5,2)
\pscircle(0.5,1){0.1}%1
\qline(0.7,1)(1.2,1)\pscircle(1.4,1){0.1}%2
\qline(1.6,1)(2.1,1)\psline[linestyle=dotted,dotsep=0.1,linewidth=0.05](2.2,1)(2.7,1)%
\qline(2.8,1)(3.4,1)\pscircle(3.6,1){0,1}%3
\psline[doubleline=true,doublesep=0.03,arrows=->,arrowlength=0.6,arrowinset=0.55](3.8,1)(4.3,1)\pscircle(4.5,1){0.1}%4
\put(0.4,1.2){\f{\lambda_1}}%1
\put(1.3,1.2){\f{\lambda_2}}%2
\put(3.5,1.2){\f{\lambda_{r-1}}}%3
\put(4.4,1.2){\f{\lambda_r}}%4
\end{pspicture}
&
\begin{pspicture}(0,0)(3,2.5)
\put(0,2.1){\ft{\textbf{D\textit{I}}:$\frac{SO(p,q)}{SO(p)\times SO(q)},\,p+q$ even}}%
\put(0,1.5){\f{l=min(p,q);}}%
\put(0,1.0){\ft{If $2\leq l \leq r-2$}}%
\end{pspicture}
\\
\begin{pspicture}(0,0)(6,2)
\pscircle(0.5,1){0.1}%1
\qline(0.7,1)(1.2,1)\psline[linestyle=dotted,dotsep=0.1,linewidth=0.05](1.3,1)(1.7,1)%
\qline(1.8,1)(2.3,1)\pscircle(2.5,1){0.1}% l-1
\qline(2.64,1.14)(2.99,1.49)\pscircle(3.13,1.63){0.1}% l
\qline(2.64,0.86)(2.99,0.51)\pscircle(3.13,0.37){0.1}% l+1
\psline[linearc=2]{<->}(3.23,1.53)(3.3,1)(3.23,0.47)%
\put(0.4,1.2){\f{\alpha_1}}\put(1.9,1.25){\f{\alpha_{l-1}}}\put(3.33,1.63){\f{\alpha_l}}\put(3.33,0.35){\f{\alpha_{l+1}}}%
\end{pspicture}
&
\begin{pspicture}(0,0)(5,2)
\pscircle(0.5,1){0.1}%1
\qline(0.7,1)(1.2,1)\pscircle(1.4,1){0.1}%2
\qline(1.6,1)(2.1,1)\psline[linestyle=dotted,dotsep=0.1,linewidth=0.05](2.2,1)(2.7,1)%
\qline(2.8,1)(3.4,1)\pscircle(3.6,1){0,1}%3
\psline[doubleline=true,doublesep=0.03,arrows=->,arrowlength=0.6,arrowinset=0.55](3.8,1)(4.3,1)\pscircle(4.5,1){0.1}%4
\put(0.4,1.2){\f{\lambda_1}}%1
\put(1.3,1.2){\f{\lambda_2}}%2
\put(3.5,1.2){\f{\lambda_{r-1}}}%3
\put(4.4,1.2){\f{\lambda_r}}%4
\end{pspicture}
&
\begin{pspicture}(0,0)(2,3)
\put(0,1){\ft{If $r=l+1$.}}
\end{pspicture}
\\
\begin{pspicture}(0,0)(6,2)
\pscircle(0.5,1){0.1}%1
\qline(0.7,1)(1.2,1)\psline[linestyle=dotted,dotsep=0.1,linewidth=0.05](1.3,1)(1.7,1)%
\qline(1.8,1)(2.3,1)\pscircle(2.5,1){0.1}% l-1
\qline(2.64,1.14)(2.99,1.49)\pscircle(3.13,1.63){0.1}% l
\qline(2.64,0.86)(2.99,0.51)\pscircle(3.13,0.37){0.1}% l+1
\put(0.4,1.2){\f{\alpha_1}}\put(1.9,1.25){\f{\alpha_{l-2}}}\put(3.33,1.63){\f{\alpha_{l-1}}}\put(3.33,0.35){\f{\alpha_{l}}}%
\end{pspicture}
&
\begin{pspicture}(0,0)(6,2)
\pscircle(0.5,1){0.1}%1
\qline(0.7,1)(1.2,1)\psline[linestyle=dotted,dotsep=0.1,linewidth=0.05](1.3,1)(1.7,1)%
\qline(1.8,1)(2.3,1)\pscircle(2.5,1){0.1}% l-1
\qline(2.64,1.14)(2.99,1.49)\pscircle(3.13,1.63){0.1}% l
\qline(2.64,0.86)(2.99,0.51)\pscircle(3.13,0.37){0.1}% l+1
\put(0.4,1.2){\f{\lambda_1}}\put(1.9,1.25){\f{\lambda_{l-2}}}\put(3.33,1.63){\f{\lambda_{l-1}}}\put(3.33,0.35){\f{\lambda_{l}}}%
\end{pspicture}
&
\begin{pspicture}(0,0)(2,3)
\put(0,1){\ft{If $r=l$.}}
\end{pspicture}
\\
\begin{pspicture}(-0.9,0)(5.1,2)
\pscircle(-0.4,1){0.1}\qline(-0.2,1)(0.3,1)%
\pscircle*(0.5,1){0.1}%1
\qline(0.7,1)(1.2,1)\psline[linestyle=dotted,dotsep=0.1,linewidth=0.05](1.3,1)(1.7,1)%
\qline(1.8,1)(2.3,1)\pscircle*(2.5,1){0.1}% l-1
\qline(2.64,1.14)(2.99,1.49)\pscircle*(3.13,1.63){0.1}% l
\qline(2.64,0.86)(2.99,0.51)\pscircle*(3.13,0.37){0.1}% l+1
\put(-0.5,1.2){\f{\alpha_1}}%
\end{pspicture}
&
\begin{pspicture}(-1,-1)(1,1)
\pscircle(0,0){0.1}\put(-0.1,0.2){\f{\lambda_1}}%
\end{pspicture}
&
\begin{pspicture}(0,0)(3,2)
\put(0,1.4){\ft{\textbf{D\textit{II}}:$\frac{SO(2r-1,1)}{SO(2r-1)}$}}\put(0,0.7){\f{l=1}}%
\end{pspicture}
\\
\begin{pspicture}(0,0)(6,2)
\pscircle*(0.5,1){0.1}%1
\qline(0.7,1)(1.2,1)\pscircle(1.4,1){0.1}%2
\qline(1.6,1)(2.1,1)\pscircle*(2.3,1){0.1}%3
\qline(2.5,1)(3.0,1)\psline[linestyle=dotted,dotsep=0.1,linewidth=0.05](3.1,1)(3.6,1)%
\qline(3.7,1)(4.2,1)\pscircle(4.4,1){0.1}%2r-2
\qline(4.54,1.14)(4.89,1.49)\qline(4.54,0.86)(4.89,0.51)%
\pscircle*(5.03,1.63){0.1}\pscircle(5.03,0.37){0.1}%
\put(0.4,1.2){\f{\alpha_1}}%1
\put(1.3,1.2){\f{\alpha_2}}%2
\put(3.7,1.25){\f{\alpha_{r-2}}}%r-2
\put(5.13,1.70){\f{\alpha_{r-1}}}\put(5.13,0.3){\f{\alpha_r}}%
\end{pspicture}
&
\begin{pspicture}(0,0)(5,2)
\pscircle(0.5,1){0.1}%1
\qline(0.7,1)(1.2,1)\pscircle(1.4,1){0.1}%2
\qline(1.6,1)(2.1,1)\psline[linestyle=dotted,dotsep=0.1,linewidth=0.05](2.2,1)(2.7,1)%
\qline(2.8,1)(3.4,1)\pscircle(3.6,1){0,1}%3
\psline[doubleline=true,doublesep=0.03,arrows=<-,arrowlength=0.6,arrowinset=0.55](3.8,1)(4.3,1)\pscircle(4.5,1){0.1}%4
\put(0.4,1.2){\f{\lambda_2}}%1
\put(1.3,1.2){\f{\lambda_4}}%2
\put(4.4,1.2){\f{\lambda_{2l}}}%4
\end{pspicture}
&
\begin{pspicture}(0,0)(3,2.5)
\put(0,2.1){\ft{\textbf{D\textit{III}}:$\frac{SO^*(2n)}{U(n)}$}}%
\put(0,1.5){\f{l=[n/2]}}%
\put(0,1.0){\ft{If $r=2l$}}%
\end{pspicture}
\\
\begin{pspicture}(0,0)(6,2)
\pscircle*(0.5,1){0.1}%1
\qline(0.7,1)(1.2,1)\pscircle(1.4,1){0.1}%2
\qline(1.6,1)(2.1,1)\pscircle*(2.3,1){0.1}%3
\qline(2.5,1)(3.0,1)\psline[linestyle=dotted,dotsep=0.1,linewidth=0.05](3.1,1)(3.6,1)%
\qline(3.7,1)(4.2,1)\pscircle(4.4,1){0.1}%2r-2
\qline(4.54,1.14)(4.89,1.49)\qline(4.54,0.86)(4.89,0.51)%
\pscircle(5.03,1.63){0.1}\pscircle(5.03,0.37){0.1}%
\psline[linearc=1]{<->}(5.17,1.49)(5.3,1)(5.17,0.55)%
\put(0.4,1.2){\f{\alpha_1}}%1
\put(1.3,1.2){\f{\alpha_2}}%2
\put(3.7,1.25){\f{\alpha_{r-2}}}%r-2
\put(5.13,1.70){\f{\alpha_{r-1}}}\put(5.13,0.3){\f{\alpha_r}}%
\end{pspicture}
&
\begin{pspicture}(0,0)(5,2)
\pscircle(0.5,1){0.1}%1
\qline(0.7,1)(1.2,1)\pscircle(1.4,1){0.1}%2
\qline(1.6,1)(2.1,1)\psline[linestyle=dotted,dotsep=0.1,linewidth=0.05](2.2,1)(2.7,1)%
\qline(2.8,1)(3.4,1)\pscircle(3.6,1){0,1}%3
\psline[doubleline=true,doublesep=0.03,arrows=->,arrowlength=0.6,arrowinset=0.55](3.8,1)(4.3,1)\pscircle(4.5,1){0.1}%4
\put(0.4,1.2){\f{\lambda_2}}%1
\put(1.3,1.2){\f{\lambda_4}}%2
\put(4.4,1.2){\f{\lambda_{2l}}}%4
\end{pspicture}
&
\begin{pspicture}(0,0)(3,2)
\put(0,1){\ft{If $r=2l+1$}}%
\end{pspicture}
\\
\begin{pspicture}(0,0)(6,2)
\pscircle(.5,1){0.1}%1
\qline(.7,1)(1.2,1)\pscircle(1.4,1){0.1}%2
\qline(1.6,1)(2.1,1)\pscircle(2.3,1){0.1}%3
\qline(2.5,1)(3.0,1)\pscircle(3.2,1){0.1}%4
\qline(3.4,1)(3.9,1)\pscircle(4.1,1){0.1}%5
\qline(2.3,1.2)(2.3,1.7)\pscircle(2.3,1.9){0.1}%6
\end{pspicture}
&
\begin{pspicture}(0,0)(6,2)
\pscircle(.5,1){0.1}%1
\qline(.7,1)(1.2,1)\pscircle(1.4,1){0.1}%2
\qline(1.6,1)(2.1,1)\pscircle(2.3,1){0.1}%3
\qline(2.5,1)(3.0,1)\pscircle(3.2,1){0.1}%4
\qline(3.4,1)(3.9,1)\pscircle(4.1,1){0.1}%5
\qline(2.3,1.2)(2.3,1.7)\pscircle(2.3,1.9){0.1}%6
\end{pspicture}
&
\begin{pspicture}(0,0)(3,2)
\put(0,1.4){\ft{\textbf{E\textit{I}}:$E_{6(6)}/Sp(4)$.}}%
\put(0,0.7){\f{l=r=6}}
\end{pspicture}
\\
\begin{pspicture}(0,0)(6,2.2)
\pscircle(.5,1){0.1}%1
\qline(.7,1)(1.2,1)\pscircle(1.4,1){0.1}%2
\qline(1.6,1)(2.1,1)\pscircle(2.3,1){0.1}%3
\qline(2.5,1)(3.0,1)\pscircle(3.2,1){0.1}%4
\qline(3.4,1)(3.9,1)\pscircle(4.1,1){0.1}%5
\qline(2.3,1.2)(2.3,1.7)\pscircle(2.3,1.9){0.1}%6
\psline[linearc=2]{<->}(1.55,0.85)(2.3,0.6)(3.05,0.85)%
\psline[linearc=3]{<->}(0.6,0.85)(2.3,0.4)(4.0,0.85)%
\put(0.4,1.2){\f{\alpha_6}}\put(1.3,1.2){\f{\alpha_5}}\put(2.4,1.2){\f{\alpha_4}}%
\put(3.1,1.2){\f{\alpha_3}}\put(4.0,1.2){\f{\alpha_1}}\put(2.5,1.8){\f{\alpha_2}}%
\end{pspicture}
&
\begin{pspicture}(0,0)(6,2)
\pscircle(.5,1){0.1}%1
\qline(.7,1)(1.2,1)\pscircle(1.4,1){0.1}%2
\psline[doubleline=true,doublesep=0.03,arrows=->,arrowlength=0.6,arrowinset=0.55](1.6,1)(2.1,1)%
\pscircle(2.3,1){0.1}\qline(2.5,1)(3.0,1)\pscircle(3.2,1){0.1}%
\put(0.4,1.2){\f{\lambda_2}}\put(1.3,1.2){\f{\lambda_4}}%
\put(2.2,1.2){\f{\lambda_3}}\put(3.1,1.2){\f{\lambda_1}}%
\end{pspicture}
&
\begin{pspicture}(0,0)(3,2)
\put(0,1.4){\ft{\textbf{E\textit{II}}:$\frac{E_{6(2)}}{SU(6)\times SU(2)}$.}}%
\put(0,0.7){\f{l=4}}
\end{pspicture}
\\
\begin{pspicture}(0,0)(6,2.2)
\pscircle(.5,1){0.1}%1
\qline(.7,1)(1.2,1)\pscircle*(1.4,1){0.1}%2
\qline(1.6,1)(2.1,1)\pscircle*(2.3,1){0.1}%3
\qline(2.5,1)(3.0,1)\pscircle*(3.2,1){0.1}%4
\qline(3.4,1)(3.9,1)\pscircle(4.1,1){0.1}%5
\qline(2.3,1.2)(2.3,1.7)\pscircle(2.3,1.9){0.1}%6
\psline[linearc=2]{<->}(0.6,0.85)(2.3,0.6)(4.0,0.85)%
\put(0.4,1.2){\f{\alpha_6}}\put(1.3,1.2){\f{\alpha_5}}\put(2.4,1.2){\f{\alpha_4}}%
\put(3.1,1.2){\f{\alpha_3}}\put(4.0,1.2){\f{\alpha_1}}\put(2.5,1.8){\f{\alpha_2}}%
\end{pspicture}
&
\begin{pspicture}(0,0)(6,2)
\pscircle(.5,1){0.1}%1
\psline[doubleline=true,doublesep=0.03,arrows=->,arrowlength=0.6,arrowinset=0.55](0.7,1)(1.2,1)%
\pscircle(1.4,1){0.1}%
\put(0.4,1.2){\f{\lambda_2}}\put(1.3,1.2){\f{\lambda_1}}%
\end{pspicture}
&
\begin{pspicture}(0,0)(3,2)
\put(0,1.4){\ft{\textbf{E\textit{III}}:$\frac{E_{6(-14)}}{SO(10)\times U(1)}$.}}%
\put(0,0.7){\f{l=2}}
\end{pspicture}
\\
\begin{pspicture}(0,0)(6,2.2)
\pscircle(.5,1){0.1}%1
\qline(.7,1)(1.2,1)\pscircle*(1.4,1){0.1}%2
\qline(1.6,1)(2.1,1)\pscircle*(2.3,1){0.1}%3
\qline(2.5,1)(3.0,1)\pscircle*(3.2,1){0.1}%4
\qline(3.4,1)(3.9,1)\pscircle(4.1,1){0.1}%5
\qline(2.3,1.2)(2.3,1.7)\pscircle*(2.3,1.9){0.1}%6
\put(0.4,1.2){\f{\alpha_6}}%
\put(4.0,1.2){\f{\alpha_1}}%
\end{pspicture}
&
\begin{pspicture}(0,0)(6,2)
\pscircle(.5,1){0.1}%1
\qline(0.7,1)(1.2,1)%
\pscircle(1.4,1){0.1}%
\put(0.4,1.2){\f{\lambda_1}}\put(1.3,1.2){\f{\lambda_6}}%
\end{pspicture}
&
\begin{pspicture}(0,0)(3,2)
\put(0,1.4){\ft{\textbf{E\textit{IV}}:$\frac{E_{6(-24)}}{F_4}$.}}%
\put(0,0.7){\f{l=2}}
\end{pspicture}
\\
\begin{pspicture}(-0.9,0)(5.1,2.1)
\pscircle(-0.4,1){0.1}%1
\qline(-0.2,1)(0.3,1)\pscircle(.5,1){0.1}%2
\qline(.7,1)(1.2,1)\pscircle(1.4,1){0.1}%3
\qline(1.6,1)(2.1,1)\pscircle(2.3,1){0.1}%4
\qline(2.5,1)(3.0,1)\pscircle(3.2,1){0.1}%5
\qline(3.4,1)(3.9,1)\pscircle(4.1,1){0.1}%6
\qline(2.3,1.2)(2.3,1.7)\pscircle(2.3,1.9){0.1}%7
\end{pspicture}
&
\begin{pspicture}(-0.9,0)(5.1,2.1)
\pscircle(-0.4,1){0.1}%1
\qline(-0.2,1)(0.3,1)\pscircle(.5,1){0.1}%2
\qline(.7,1)(1.2,1)\pscircle(1.4,1){0.1}%3
\qline(1.6,1)(2.1,1)\pscircle(2.3,1){0.1}%4
\qline(2.5,1)(3.0,1)\pscircle(3.2,1){0.1}%5
\qline(3.4,1)(3.9,1)\pscircle(4.1,1){0.1}%6
\qline(2.3,1.2)(2.3,1.7)\pscircle(2.3,1.9){0.1}%7
\end{pspicture}
&
\begin{pspicture}(0,0)(3,2)
\put(0,1.4){\ft{\textbf{E\textit{V}}:$E_{7(7)}/SU(8)$.}}%
\put(0,0.7){\f{l=r=7}}
\end{pspicture}
\\
\begin{pspicture}(-0.9,0)(5.1,2.2)
\pscircle*(-0.4,1){0.1}%
\qline(-0.2,1)(0.3,1)\pscircle(.5,1){0.1}%2
\qline(.7,1)(1.2,1)\pscircle*(1.4,1){0.1}%3
\qline(1.6,1)(2.1,1)\pscircle(2.3,1){0.1}%4
\qline(2.5,1)(3.0,1)\pscircle(3.2,1){0.1}%5
\qline(3.4,1)(3.9,1)\pscircle(4.1,1){0.1}%6
\qline(2.3,1.2)(2.3,1.7)\pscircle*(2.3,1.9){0.1}%7
\put(-0.5,1.2){\f{\alpha_7}}%
\put(0.4,1.2){\f{\alpha_6}}\put(1.3,1.2){\f{\alpha_5}}\put(2.4,1.2){\f{\alpha_4}}%
\put(3.1,1.2){\f{\alpha_3}}\put(4.0,1.2){\f{\alpha_1}}\put(2.5,1.8){\f{\alpha_2}}%
\end{pspicture}
&
\begin{pspicture}(0,0)(6,2)
\pscircle(.5,1){0.1}%1
\qline(.7,1)(1.2,1)\pscircle(1.4,1){0.1}%2
\psline[doubleline=true,doublesep=0.03,arrows=->,arrowlength=0.6,arrowinset=0.55](1.6,1)(2.1,1)%
\pscircle(2.3,1){0.1}\qline(2.5,1)(3.0,1)\pscircle(3.2,1){0.1}%
\put(0.4,1.2){\f{\lambda_1}}\put(1.3,1.2){\f{\lambda_3}}%
\put(2.2,1.2){\f{\lambda_4}}\put(3.1,1.2){\f{\lambda_6}}%
\end{pspicture}
&
\begin{pspicture}(0,0)(3,2)
\put(0,1.4){\ft{\textbf{E\textit{VI}}:$\frac{E_{7(-5)}}{SO(12)\times SU(2)}$.}}%
\put(0,0.7){\f{l=4}}
\end{pspicture}
\\
\begin{pspicture}(-0.9,0)(5.1,2.2)
\pscircle(-0.4,1){0.1}%
\qline(-0.2,1)(0.3,1)\pscircle(.5,1){0.1}%2
\qline(.7,1)(1.2,1)\pscircle*(1.4,1){0.1}%3
\qline(1.6,1)(2.1,1)\pscircle*(2.3,1){0.1}%4
\qline(2.5,1)(3.0,1)\pscircle*(3.2,1){0.1}%5
\qline(3.4,1)(3.9,1)\pscircle(4.1,1){0.1}%6
\qline(2.3,1.2)(2.3,1.7)\pscircle*(2.3,1.9){0.1}%7
\put(-0.5,1.2){\f{\alpha_7}}%
\put(0.4,1.2){\f{\alpha_6}}\put(1.3,1.2){\f{\alpha_5}}\put(2.4,1.2){\f{\alpha_4}}%
\put(3.1,1.2){\f{\alpha_3}}\put(4.0,1.2){\f{\alpha_1}}\put(2.5,1.8){\f{\alpha_2}}%
\end{pspicture}
&
\begin{pspicture}(0,0)(6,2)
\pscircle(.5,1){0.1}%1
\qline(.7,1)(1.2,1)\pscircle(1.4,1){0.1}%2
\psline[doubleline=true,doublesep=0.03,arrows=<-,arrowlength=0.6,arrowinset=0.55](1.6,1)(2.1,1)%
\pscircle(2.3,1){0.1}%
\put(0.4,1.2){\f{\lambda_1}}\put(1.3,1.2){\f{\lambda_6}}%
\put(2.2,1.2){\f{\lambda_7}}%
\end{pspicture}
&
\begin{pspicture}(0,0)(3,2)
\put(0,1.4){\ft{\textbf{E\textit{VII}}:$\frac{E_{7(-25)}}{E_6\times U(1)}$.}}%
\put(0,0.7){\f{l=3}}
\end{pspicture}
\\
\begin{pspicture}(-1.6,0)(4.3,2.15)
\pscircle(-1.3,1){0.1}%
\qline(-1.1,1)(-0.6,1)\pscircle(-0.4,1){0.1}%1
\qline(-0.2,1)(0.3,1)\pscircle(.5,1){0.1}%2
\qline(.7,1)(1.2,1)\pscircle(1.4,1){0.1}%3
\qline(1.6,1)(2.1,1)\pscircle(2.3,1){0.1}%4
\qline(2.5,1)(3.0,1)\pscircle(3.2,1){0.1}%5
\qline(3.4,1)(3.9,1)\pscircle(4.1,1){0.1}%6
\qline(2.3,1.2)(2.3,1.7)\pscircle(2.3,1.9){0.1}%7
\end{pspicture}
&
\begin{pspicture}(-1.6,0)(4.3,2.15)
\pscircle(-1.3,1){0.1}%
\qline(-1.1,1)(-0.6,1)\pscircle(-0.4,1){0.1}%1
\qline(-0.2,1)(0.3,1)\pscircle(.5,1){0.1}%2
\qline(.7,1)(1.2,1)\pscircle(1.4,1){0.1}%3
\qline(1.6,1)(2.1,1)\pscircle(2.3,1){0.1}%4
\qline(2.5,1)(3.0,1)\pscircle(3.2,1){0.1}%5
\qline(3.4,1)(3.9,1)\pscircle(4.1,1){0.1}%6
\qline(2.3,1.2)(2.3,1.7)\pscircle(2.3,1.9){0.1}%7
\end{pspicture}
&
\begin{pspicture}(0,0)(3,2)
\put(0,1.4){\ft{\textbf{E\textit{VIII}}:$E_{8(8)}/SO(16)$.}}%
\put(0,0.7){\f{l=r=8}}
\end{pspicture}
\\
\begin{pspicture}(-1.6,0)(4.3,2)
\pscircle(-1.3,1){0.1}%
\qline(-1.1,1)(-0.6,1)\pscircle(-0.4,1){0.1}%
\qline(-0.2,1)(0.3,1)\pscircle(.5,1){0.1}%2
\qline(.7,1)(1.2,1)\pscircle*(1.4,1){0.1}%3
\qline(1.6,1)(2.1,1)\pscircle*(2.3,1){0.1}%4
\qline(2.5,1)(3.0,1)\pscircle*(3.2,1){0.1}%5
\qline(3.4,1)(3.9,1)\pscircle(4.1,1){0.1}%6
\qline(2.3,1.2)(2.3,1.7)\pscircle*(2.3,1.9){0.1}%7
\put(-0.5,1.2){\f{\alpha_7}}%
\put(0.4,1.2){\f{\alpha_6}}\put(1.3,1.2){\f{\alpha_5}}\put(2.4,1.2){\f{\alpha_4}}%
\put(3.1,1.2){\f{\alpha_3}}\put(4.0,1.2){\f{\alpha_1}}\put(2.5,1.8){\f{\alpha_2}}%
\put(-1.4,1.2){\f{\alpha_8}}
\end{pspicture}
&
\begin{pspicture}(0,0)(6,2)
\pscircle(.5,1){0.1}%1
\qline(.7,1)(1.2,1)\pscircle(1.4,1){0.1}%2
\psline[doubleline=true,doublesep=0.03,arrows=->,arrowlength=0.6,arrowinset=0.55](1.6,1)(2.1,1)%
\pscircle(2.3,1){0.1}\qline(2.5,1)(3.0,1)\pscircle(3.2,1){0.1}%
\put(0.4,1.2){\f{\lambda_8}}\put(1.3,1.2){\f{\lambda_7}}%
\put(2.2,1.2){\f{\lambda_6}}\put(3.1,1.2){\f{\lambda_1}}%
\end{pspicture}
&
\begin{pspicture}(0,0)(3,2)
\put(0,1.4){\ft{\textbf{E\textit{IX}}:$\frac{E_{8(-24)}}{E_7\times SU(2)}$.}}%
\put(0,0.7){\f{l=4}}
\end{pspicture}
\\
\begin{pspicture}(0,0)(6,2)
\pscircle(.5,1){0.1}%1
\qline(.7,1)(1.2,1)\pscircle(1.4,1){0.1}%2
\psline[doubleline=true,doublesep=0.03,arrows=->,arrowlength=0.6,arrowinset=0.55](1.6,1)(2.1,1)%
\pscircle(2.3,1){0.1}\qline(2.5,1)(3.0,1)\pscircle(3.2,1){0.1}%
\end{pspicture}
&
\begin{pspicture}(0,0)(6,2)
\pscircle(.5,1){0.1}%1
\qline(.7,1)(1.2,1)\pscircle(1.4,1){0.1}%2
\psline[doubleline=true,doublesep=0.03,arrows=->,arrowlength=0.6,arrowinset=0.55](1.6,1)(2.1,1)%
\pscircle(2.3,1){0.1}\qline(2.5,1)(3.0,1)\pscircle(3.2,1){0.1}%
\end{pspicture}
&
\begin{pspicture}(0,0)(3,2)
\put(0,1.4){\ft{\textbf{F\textit{I}}:$\frac{F_{4(4)}}{Sp(3)\times SU(2)}$.}}%
\put(0,0.7){\f{r=l=4}}
\end{pspicture}
\\
\begin{pspicture}(0,0)(6,2)
\pscircle*(.5,1){0.1}%1
\qline(.7,1)(1.2,1)\pscircle*(1.4,1){0.1}%2
\psline[doubleline=true,doublesep=0.03,arrows=->,arrowlength=0.6,arrowinset=0.55](1.6,1)(2.1,1)%
\pscircle*(2.3,1){0.1}\qline(2.5,1)(3.0,1)\pscircle(3.2,1){0.1}%
\put(0.4,1.2){\f{\alpha_1}}\put(1.3,1.2){\f{\alpha_2}}%
\put(2.2,1.2){\f{\alpha_3}}\put(3.1,1.2){\f{\alpha_4}}%
\end{pspicture}
&
\begin{pspicture}(-1,-1)(1,1)
\pscircle(0,0){0.1}\put(-0.1,0.2){\f{\lambda_1}}%
\end{pspicture}
&
\begin{pspicture}(0,0)(3,2)
\put(0,1.4){\ft{\textbf{F\textit{II}}:$F_{4(-20)}/SO(9)$.}}%
\put(0,0.7){\f{l=1}}
\end{pspicture}
\\
\begin{pspicture}(0,0)(6,2)
\pscircle(0.5,1){0.1}%1
\psline[arrows=->,arrowlength=0.2,arrowinset=0.15,arrowsize=11pt](0.7,1)(1.2,1)\psline(0.7,1.08)(1.12,1.08)%
\psline(0.7,0.92)(1.12,0.92)\pscircle(1.4,1){0.1}%2
\end{pspicture}
&
\begin{pspicture}(0,0)(6,2)
\pscircle(0.5,1){0.1}%1
\psline[arrows=->,arrowlength=0.2,arrowinset=0.15,arrowsize=11pt](0.7,1)(1.2,1)\psline(0.7,1.08)(1.12,1.08)%
\psline(0.7,0.92)(1.12,0.92)\pscircle(1.4,1){0.1}%2
\end{pspicture}
&
\begin{pspicture}(0,0)(3,2)
\put(0,1.4){\ft{\textbf{G}:$\frac{G_{2(2)}}{SU(2)\times SU(2)}$.}}%
\put(0,0.7){\f{l=r=2}}
\end{pspicture}
\\
\end{longtable}

\begin{longtable}{|l|ccc|}
\caption[Table 2]{Multiplicities of the restricted simple roots}\\
\hline Type & & $m_{\lambda_i}$ & $m_{2\lambda_i}$\\
\hline \endfirsthead
\caption[Table 1]{continued}\\
\hline Type & & $m_\lambda$ & $m_{2\lambda}$ \\
\hline \endhead \hline \multicolumn{4}{r}{Continued on next page}
\endfoot
\hline\endlastfoot
 \textbf{A\textit{I}} & $\forall i$& 1  &  0  \\
 \textbf{A\textit{II}} & $\forall i$&  4 &  0  \\
 \textbf{A\textit{III}}; $2\leq l\leq \tfrac{r}{2}$& $i<l$ & 2 & 0  \\
                                                    & $i=l$ & $2(r-2l+1)$ & 1\\
 \textbf{A\textit{III}}; $r=2l-1$ &  $i<l$ & 2 & 0 \\
                                  &  $i=l$ & 1 & 0 \\
 \textbf{A\textit{IV}}            &        & $2(r-1)$ & 1 \\
 \textbf{B\textit{I}}             &  $i<l$ & 1 & 0 \\
                                  &  $i=l$ & $2(r-l)+1$ & 0 \\
 \textbf{B\textit{II}}            &  $\forall i$ & $2r-1$ & 0\\
 \textbf{C\textit{I}}             &  $\forall i$ & 1      & 0 \\
 \textbf{C\textit{II}}; $1\leq l \leq \tfrac{1}{2} (r-1)$ & $i<2l$ & 4 & 0\\
                                                          & $i=2l$ & $4(r-2l)$ & 3 \\
 \textbf{C\textit{II}}; $2\leq l = \tfrac{1}{2}r$         & $i<2l$ &  4      &  0 \\
                                                          & $i=2l$ &  3      &  0  \\
 \textbf{D\textit{I}}; $2\leq l \leq r-2$                 & $i<l$  &  1      &  0 \\
                                                          & $i=l$  &  $2(r-l)$& 0   \\
 \textbf{D\textit{I}}; $l=r-1$                            & $i<l$  & 1        & 0   \\
                                                          & $i=l$  & 2        & 0\\
 \textbf{D\textit{I}}; $l=r$                              & $\forall i$ & 1 & 0 \\
 \textbf{D\textit{II}}                                    & $\forall i$ & 1 & 0  \\
 \textbf{D\textit{III}}; $r=2l$                           & $i<2l$      & 4 & 0 \\
                                                          & $i=2l$      & 1 & 0 \\
 \textbf{D\textit{III}}; $r=2l+1$                         & $i<2l$      & 4 & 1 \\
 \textbf{E\textit{I}}                                     & $\forall i$ & 1 & 0 \\
 \textbf{E\textit{II}}                                    & $i=2,4$     & 1 & 0  \\
                                                          & $i=1,3$     & 2 & 0 \\
 \textbf{E\textit{III}}                                   & $i=1$       & 8 & 1\\
                                                          & $i=2$       & 6 & 0 \\
 \textbf{E\textit{IV}}                                    & $\forall i$ & 8 & 0  \\
 \textbf{E\textit{V}}                                     & $\forall i$ & 1 & 0 \\
 \textbf{E\textit{VI}}                                    & $i=1,3$     & 1 & 0  \\
                                                          & $i=2,4$     & 4 & 0 \\
 \textbf{E\textit{VII}}                                   & $i=1,6$     & 8 & 0  \\
                                                          & $i=7$       & 1 & 0 \\
 \textbf{E\textit{VIII}}                                  & $\forall i$ & 1 & 0  \\
 \textbf{E\textit{IX}}                                    & $i=1,6$     & 8 & 0  \\
                                                          & $i=7,8$     & 1 & 0  \\
 \textbf{F\textit{I}}                                     & $\forall i$ & 1 & 0 \\
 \textbf{F\textit{II}}                                    & $\forall i$ & 8 & 7 \\
 \textbf{G\textit{I}}                                     & $\forall i$ & 1 & 0 \\
\end{longtable}

\end{appendices}

\end{document}